\documentclass[aps,prfluid,preprint,superscriptaddress,showkeys]{revtex4-2}
\usepackage{graphicx}
\usepackage{dcolumn}
\usepackage{bm}
\usepackage[mathlines]{lineno}
\usepackage{epsfig}
\usepackage{epstopdf}
\usepackage{xcolor}
\usepackage{mathtools}
\usepackage{subfig}

\begin{document}
\title{Scaling laws of two-dimensional incompressible turbulent transport}
\author{D. I. Palade}
\email{dragos.palade@inflpr.ro}
\affiliation{National Institute for Laser, Plasma and Radiation Physics,	M\u{a}gurele, Romania }
\affiliation{Faculty of Physics, University of Bucharest, Romania}
\author{L. M. Pomârjanschi}
\affiliation{National Institute for Laser, Plasma and Radiation Physics,	M\u{a}gurele, Romania }
\affiliation{Faculty of Physics, University of Bucharest, Romania}
\author{M. Ghiță}
\affiliation{Faculty of Physics, University of Bucharest, Romania}

\date{\today}

\begin{abstract}
The diffusive transport in two-dimensional incompressible turbulent fields is investigated with the aid of high-quality direct numerical simulations. Three classes of turbulence spectra that are able to capture both short and long-range time-space correlations and oscillating features are employed. We report novel scaling laws that depart from the $\gamma=7/10$ paradigm of percolative exponents and are dependent on the features of turbulence. A simple relation between diffusion in the percolative and frozen regimes is found. The importance of discerning between differential and integral characteristic scales is emphasized. 
\end{abstract}


\maketitle

\section{Introduction}
\label{section_1}

The problem of turbulence remains one of the most elusive topics of modern physics. It is of immediate interest for a variety of systems such as (geophysical) fluids \cite{RevModPhys.71.S383}, astrophysical systems \cite{Brandenburg2013}, magnetically confined plasmas \cite{doi:10.1063/1.1853385,balescu2005aspects}, magnetic field line dynamics \cite{doi:10.1063/1.2776905}, quantum systems \cite{Vinen2002,doi:10.1146/annurev-conmatphys-062910-140533,palade2018schr}, etc. While being a compact problem, it can be formally split into two complementary parts: the characterization of turbulence and the characterization of transport. 

The first part, which is a broad topic beyond the scope of the present work, is related to the representation of a turbulent field $\mathbf{v}(\mathbf{x},t)$. This is, in general, a difficult task which may require refined solutions of non-linear fluid \cite{foias2001navier} or kinetic approaches \cite{Garbet_2010,GORLER20117053}. It is often the case that the resulting field $\mathbf{v}\left(\mathbf{x},t\right)$ exhibits chaotic features that emerge, develop and saturate as turbulent states. These features can be captured via their Eulerian statistical properties: average, distribution and spectra. 

The second part is the main concern of the present paper and can be stated as it follows: \emph{Given the statistical properties  of a turbulent velocity flow $\mathbf{v}\left(\mathbf{x},t\right)$, characterize the transport coefficients (convection $\mathbf{V}(t)$ and diffusion $\hat{D}(t)$) of passive scalars advected by this field}. 

The advection is apparent at the level of a continuity-like equation $\partial_t n+\nabla \left(n\mathbf{v}\right)=0$ for a scalar field $n(\mathbf{x},t)$. In this Eulerian view, the transport coefficients $\mathbf{V}(t), \hat{D}(t)$ appear in the Fick's law for the scalar flux $\mathbf{\Gamma} = \langle n \mathbf{v}\rangle \equiv \mathbf{V}(t)n + \hat{D}(t)\nabla n$, where $\langle\cdot\rangle$ is the statistical averaging operation over realizations of turbulent flows $\mathbf{v}(\mathbf{x},t)$. From a Lagrangian perspective, one can show that, under the assumption of scale-separation \cite{balescu2005aspects}, the transport coefficients are related to statistical averages $\mathbf{V}(t)=d_t\langle \mathbf{x}(t)\rangle$, $2\hat{D}(t)=d_t\left(\langle \mathbf{x}^2(t)\rangle-\langle \mathbf{x}(t)\rangle^2\right)$ over ensembles of trajectories $\{\mathbf{x}(t)\}$. The latter are driven by the corresponding ensemble of stochastic velocity fields via $\dot{\mathbf{x}}(t)=\mathbf{v}(\mathbf{x}(t),t)$, known as a V-Langevin equation \cite{BALESCU200762,Palade2021}. 

The mean square displacement $\sigma(t) = \langle \mathbf{x}^2(t)\rangle-\langle \mathbf{x}(t)\rangle^2$, in the asymptotic limit, exhibits algebraic behaviour $\lim\limits_{t\to\infty}\sigma(t)\sim t^\alpha$. Depending on the features of turbulence, one can get $\alpha=1$ (normal, molecular-like diffusion), $\alpha=2$ (ballistic transport), $0<\alpha<1$ (sub-diffusion) or $1<\alpha$ (superdiffusion). The last two cases are coined in literature \emph{anomalous transport} \cite{balescu2005aspects,klages2008anomalous}. A sufficiently time-decorrelated turbulent flow $\mathbf{v}(\mathbf{x},t)$ can lead to normal diffusion $\alpha \to 1$, but, if consistent spatial correlations are present, the transport is still named \emph{anomalous}. This is a consequence of peculiar behaviour of the asymptotic diffusion algebraic scaling with the strength of turbulence $V=\sqrt{\langle \mathbf{v}^2(\mathbf{x},t)\rangle}$ as $D^\infty=D(t\to\infty)\sim V^\gamma$, that may be different from $\gamma = 2$ which is specific to normal (molecular) diffusion.

In the present work we investigate the scaling laws of the turbulent transport, for the special case of two-dimensional incompressible flows $\nabla \cdot \mathbf{v}(\mathbf{x},t)=0$. The latter is a paradigm for the $\mathbf{E}\times\mathbf{B}$ drift motion of charged particles in magnetically confined plasmas and for the magnetic field lines wandering in astrophysical turbulence. Both classes of systems are notorious for their ability to manifest anomalous transport \cite{doi:10.1063/1.859358,doi:10.1063/1.871453}. 

A lot of effort has been directed towards this problem in the past decades, from Bohm's diffusion \cite{bohm1949characteristics}, to Corssin's approximation \cite{CORRSIN1959161}, Kraichnan analysis \cite{RHKraichnan_1980}, Isichenko's pioneering work \cite{RevModPhys.64.961}, numerical experiments \cite{M.Ottaviani1992,Hauff_2010,doi:10.1063/1.2360173,PhysRevE.54.1857,PhysRevLett.76.4360,PhysRevE.63.066405} and other modern, theoretical, approaches \cite{PhysRevE.58.7359,doi:10.1063/1.2776905} etc. Despite such extensive endeavors, the problem is not solved yet: we do not have simple relations between the Eulerian (measurable) characteristics of a turbulent flow and the transport coefficients in all physical regimes of interest. The most promising analytical result was due to Isichenko \cite{RevModPhys.64.961} who showed, with the aid of percolation theory, that, in the strong turbulence (percolative) regime $\gamma = 7/10$, in contrast with Bohm's \cite{bohm1949characteristics} $\gamma=1$ value. The universality of this value has been put to the test in the following years by semi-analytical approximations \cite{PhysRevE.58.7359,doi:10.1063/1.2776905}, which did not confirm it. On the numerical front \cite{M.Ottaviani1992,Hauff_2010,PhysRevE.54.1857,PCastiglione_2000} many investigations have found exponents close to the analytical $\gamma = 7/10$ value. Unfortunately, the results are widely scattered, with relatively large uncertainties, poor numerical resolution and only for limited classes of turbulence spectra. 

For all these reasons, in the present work we intend to use high-quality direct numerical simulations \cite{Palade2021} to evaluate the turbulent transport in two-dimensional incompressible fields. In particular, we search for scaling laws of diffusive processes and their relation with statistical spectra/correlations. The universality of Isichenko's result is put to the test. For this analysis, we employ three classes of spectra that are able to capture both short and long-range correlations in space and time or the existence of oscillating behaviour. 

The paper is structured as it follows. In the Theory section \eqref{section_2} we present the statistical description of turbulence, the associated families of spectra/correlations to be investigated and the approach of direct numerical simulations (DNS). In the Result section \eqref{section_3} we analyze the numerical results on diffusion and scaling exponents. The final section \eqref{section_4} is devoted to conclusions and perspectives.

\section{Theory}
\label{section_2}

We consider a two-dimensional incompressible turbulent flow $\nabla\cdot\mathbf{v}(\mathbf{x},t)=0$ that can be expressed in terms of a scalar field (stream-function) $\phi(\mathbf{x},t)$ as $\mathbf{v}(\mathbf{x},t) = \hat{e}_z\times \nabla\phi(\mathbf{x},t)$ with $\mathbf{x}\equiv (x,y)$. In fusion plasmas \cite{PhysRevE.58.7359} the stream-function is related to the electrostatic potential while in astrophysical magnetic turbulence to components of the magnetic vector potential \cite{Hauff_2010}. A statistical ensemble of potentials $\{\phi(\mathbf{x},t)\}$ which are homogeneous Gaussian random fields \cite{Palade2021} with zero average $\langle \phi(\mathbf{x},t)\rangle = 0$ and known Eulerian correlation $\mathcal{E}(|\mathbf{x}-\mathbf{x}^\prime|,|t-t^\prime|) = \langle \phi(\mathbf{x},t)\phi(\mathbf{x}^\prime,t^\prime)\rangle$ drives an associated ensemble of trajectories $\{\mathbf{x}(t)\}$ via the V-Langevin eq:

\begin{align}
	\label{eq_1}
	\frac{d\mathbf{x}(t)}{dt}=\mathbf{v}(\mathbf{x}(t),t) = \hat{e}_z\times \nabla\phi(\mathbf{x}(t),t), \hspace{1cm} \mathbf{x}(0)=0 .
\end{align}

The resulting trajectories $\{\mathbf{x}(t)\}$ are used to evaluate the running diffusion coefficients as  $D_x(t) = 1/2d_t\langle x^2(t)\rangle$. The homogeneous correlation $\mathcal{E}(|\mathbf{x}-\mathbf{x}^\prime|,|t-t^\prime|)$ is related to the turbulence spectrum $S(\mathbf{k},\omega) = \langle |\tilde\phi(\mathbf{k},\omega)|^2\rangle$ via a Fourier transform. 
We shall assume Gaussian turbulence, i.e. a normal distribution for the values of the potential $\phi(\mathbf{x},t)$. The implications of non-Gaussianity have been discussed elsewhere \cite{palade_pomarjanschi_2022} and found to be small.

\subsection{Correlation functions and associated spectra}
\label{section_2.1}

While distinct physical systems or dynamical regimes can lead to different turbulence spectra, there are similar features shared among most cases of interest. The spectrum $S(\mathbf{k},\omega)$ is usually peaked at or near the $|\mathbf{k}| =0 ,\omega =0$ values and decays in the asymptotic limit $|\mathbf{k}|,\omega\to\infty$, at least algebraically \cite{Zweben_1985,Levinson_1984, PhysRevLett.70.3736,10.1088/1741-4326/acb959}. In order to encompass such gross properties, we consider here three classes of parametric turbulent fields as separable, isotropic, correlation functions $\mathcal{E}_{\nu/\mu}^{(j)}(\mathbf{x},t)$ and their corresponding spectra $S_{\nu/\mu}^{(j)}(\mathbf{k},\omega)$.

\begin{figure}
	\subfloat[\label{fig_1a}]{
		\includegraphics[width=.48\linewidth]{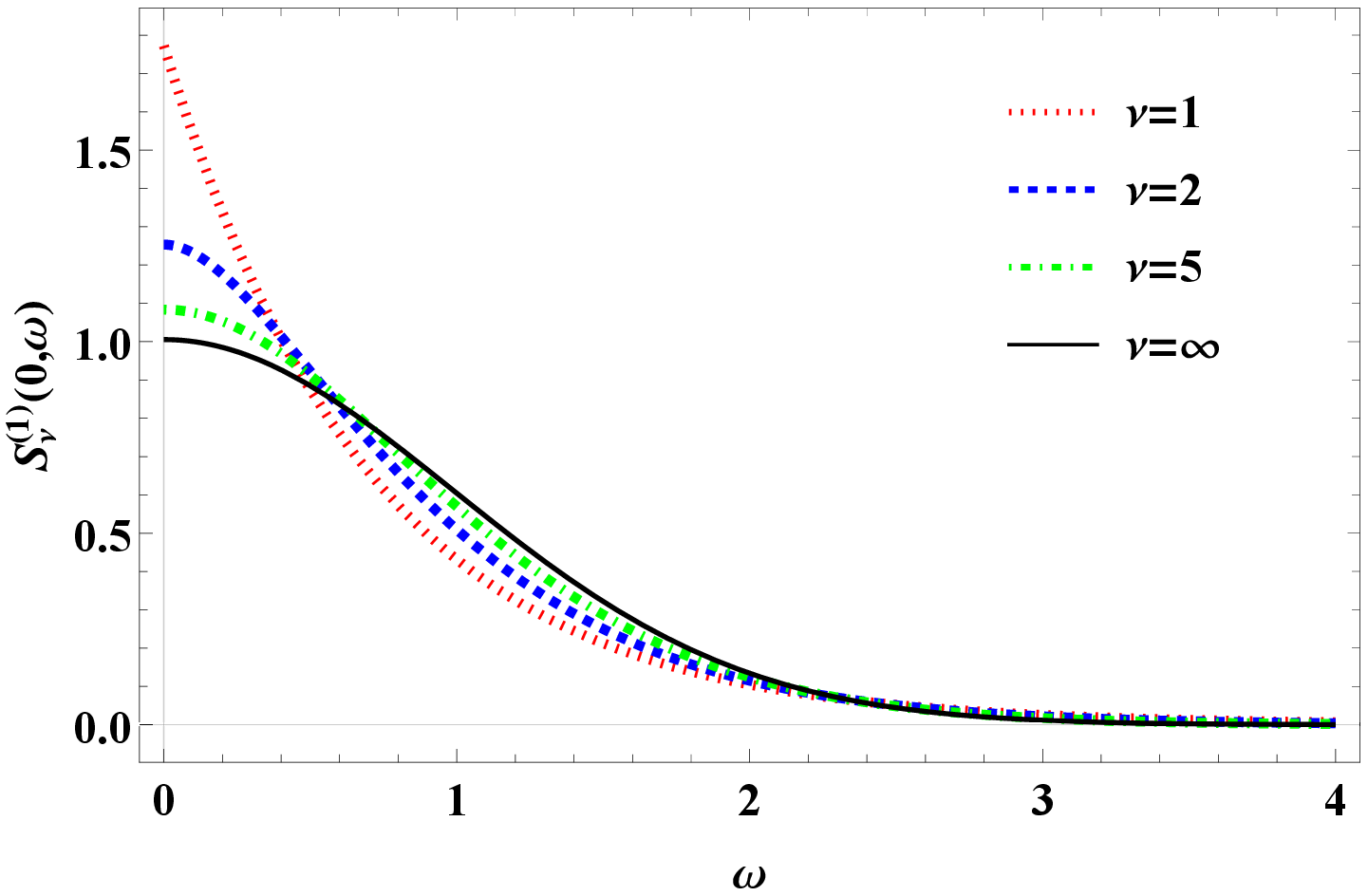}}	\hspace*{0.1cm}
	\subfloat[\label{fig_1b}]{
		\includegraphics[width=.48\linewidth]{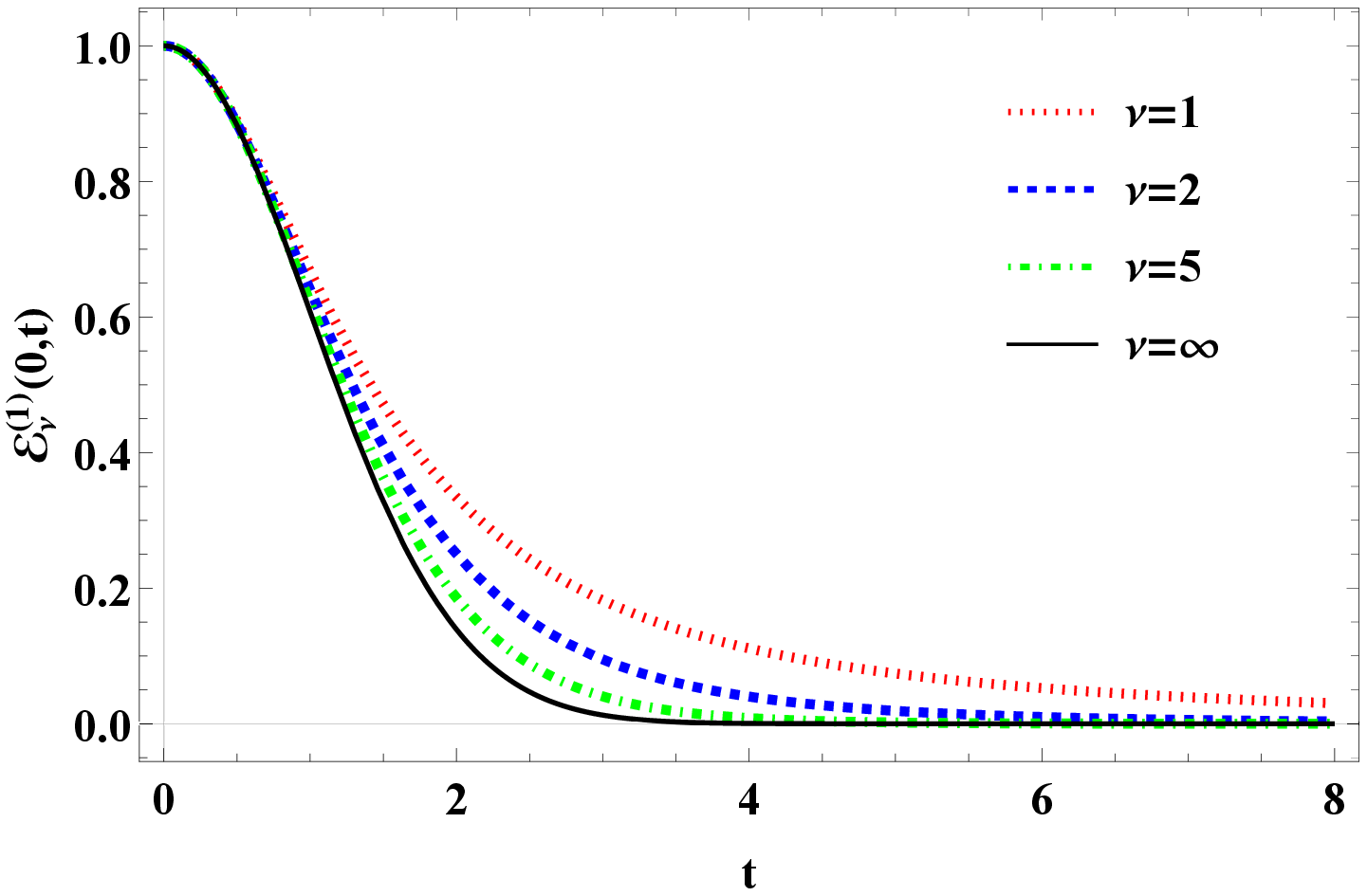}}
	\caption{Spectra $S_{\nu}^{(1)}(\mathbf{0},\omega)$ and associated correlation functions $\mathcal{E}_{\nu}^{(1)}(\mathbf{0},t)$ for several values of the parameter $\nu$ and $\tau_c=1,\lambda_c=1$. The black lines correspond to the Gaussian limit.}
\end{figure}

The correlation $\mathcal{E}_{\nu}^{(1)}(\mathbf{x},t)$ (eq. \eqref{eq_2}) consists of a Gaussian space profile with an algebraic time decay. In the limit $\nu\to 1$, the time dependence is a long-range Lorentzian function, while for $\nu\to\infty$ it becomes a short-ranged Gaussian. $K_a(z)$ is the modified Bessel function of the second kind. In Figs. \eqref{fig_1a}-\eqref{fig_1b} we plot the correlation $\mathcal{E}_{\nu}^{(1)}(\mathbf{0},t)$ and the spectrum $S_{\nu}^{(1)}(\mathbf{0},\omega)$ for different parametric values $\nu$, fixing $\tau_c=1$.

\begin{equation}
	\label{eq_2}
	S_{\nu}^{(1)}(\mathbf{k},\omega) \sim \left| \omega\tau_c \right| ^{\nu -\frac{1}{2}} K_{\nu -\frac{1}{2}}\left(\sqrt{2\nu } \left| \omega \right|\tau_c \right)e^{-\frac{\mathbf{k}^2\lambda_c^2}{2}}\longrightarrow \mathcal{E}_{\nu}^{(1)}(\mathbf{x},t)= e^{-\frac{\mathbf{x}^2}{2\lambda_c^2}}(1+\frac{t^2}{2\tau_c^2\nu})^{-\nu}.\end{equation}

\begin{figure}
	\subfloat[\label{fig_1c}]{
		\includegraphics[width=.48\linewidth]{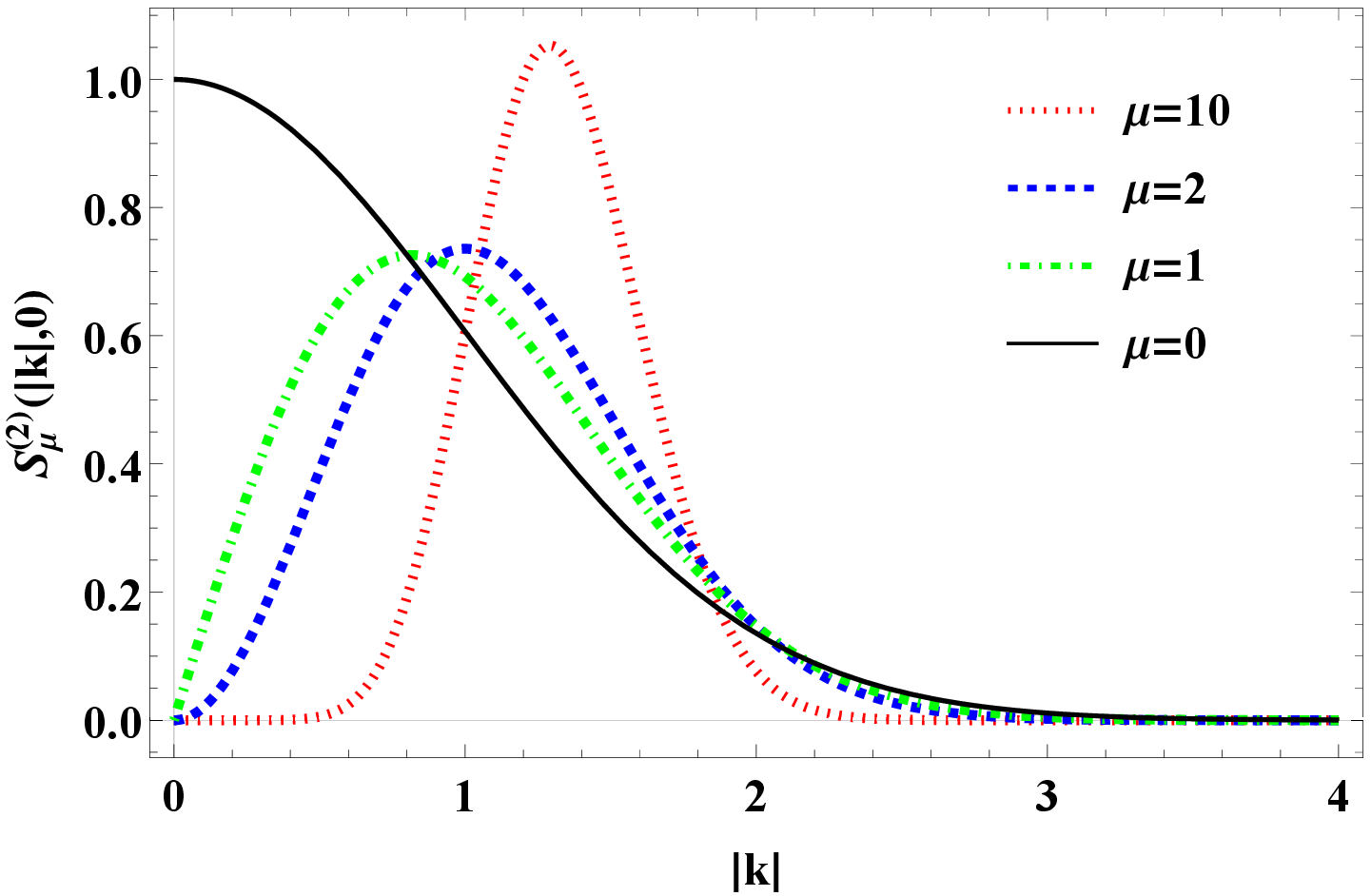}%
	}	\hspace*{0.1cm}
	\subfloat[\label{fig_1d}]{
		\includegraphics[width=.48\linewidth]{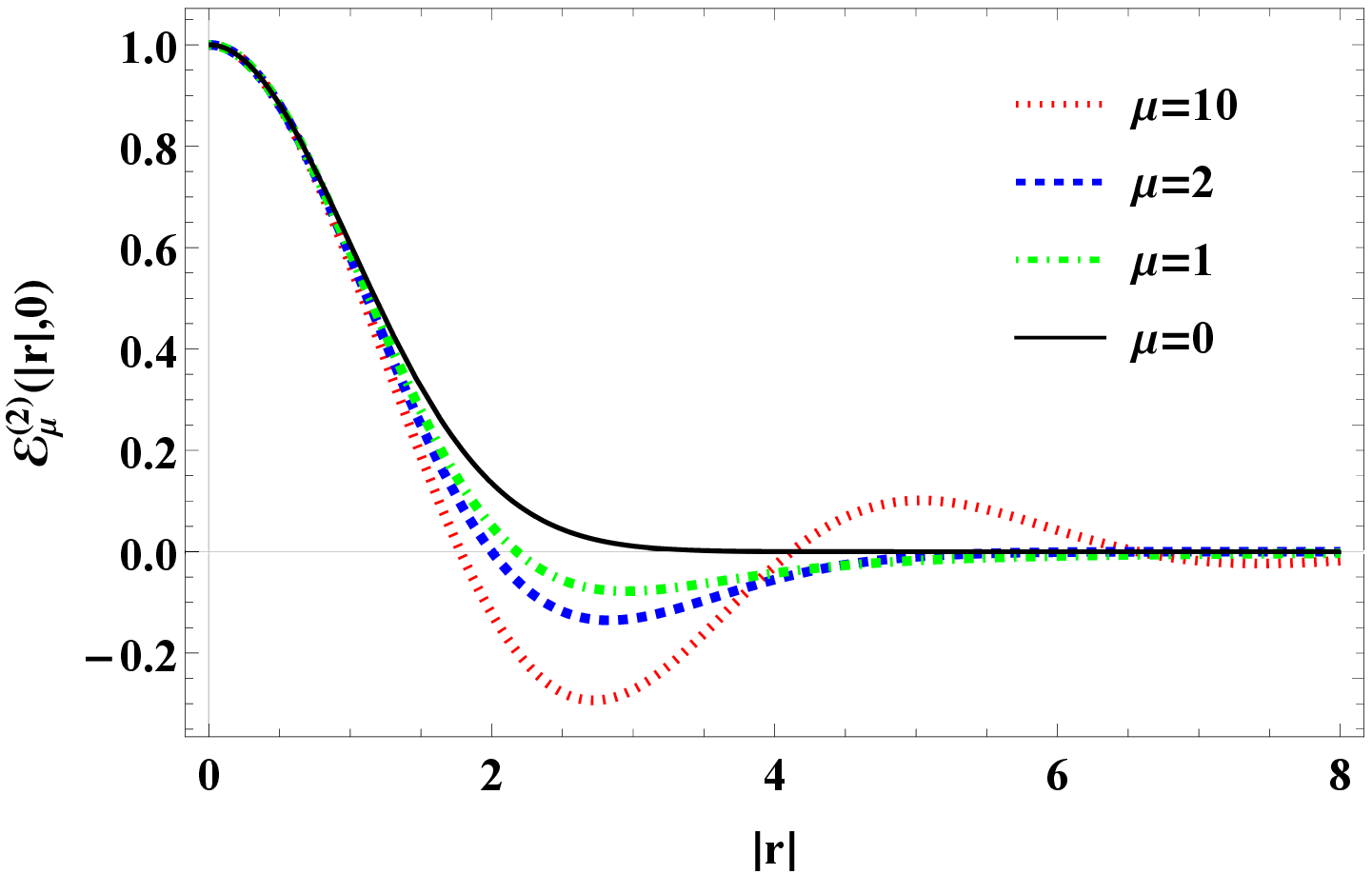}%
	}
	\caption{Spectra $S_{\mu}^{(2)}(|\mathbf{k}|,0)$ and the associated correlation functions $\mathcal{E}_{\mu}^{(2)}(|\mathbf{r}|,0)$ for several values of the parameter $\mu$ and $\tau_c=1,\lambda_c=1$. The black lines correspond to the Gaussian limit.}
\end{figure}

The correlation $\mathcal{E}_{\mu}^{(2)}(\mathbf{x},t)$ eq. \eqref{eq_3} exhibits space oscillations and it is associated to a spectrum dominated by an energy-range, thus, peaked at a non-zero $|\mathbf{k}|$ value:
\begin{equation}
	\label{eq_3}
S_{\mu}^{(2)}(\mathbf{k},\omega) \sim \left| \mathbf{k}\lambda_c \right| ^{\mu} e^{-\left| \mathbf{k}\lambda_c \right|^2(2+\mu)/4}e^{-\frac{\omega^2\tau_c^2}{2}}\longrightarrow \mathcal{E}_{\mu}^{(2)}(\mathbf{x},t)= e^{-\frac{t^2}{2\tau_c^2}}L_{-\frac{\mu }{2}-1}\left(-\frac{|\mathbf{x}|^2}{2 \left(\frac{\mu }{2}+1\right) \lambda_c^2}\right)\end{equation}
where $L_a(x)$ is the Laguerre polynomial. In Figs. \eqref{fig_1c}-\eqref{fig_1d} we plot the correlation $\mathcal{E}_{\mu}^{(2)}(|\mathbf{r}|,0)$ and the spectrum $S_{\mu}^{(2)}(|\mathbf{k}|,0)$. The parameter $\mu$ controls space oscillations of $\mathcal{E}^{(2)}$ and, in the limit $\mu=0$, is equivalent with a Gaussian correlation. 
\begin{figure}
	\subfloat[\label{fig_1e}]{
		\includegraphics[width=.48\linewidth]{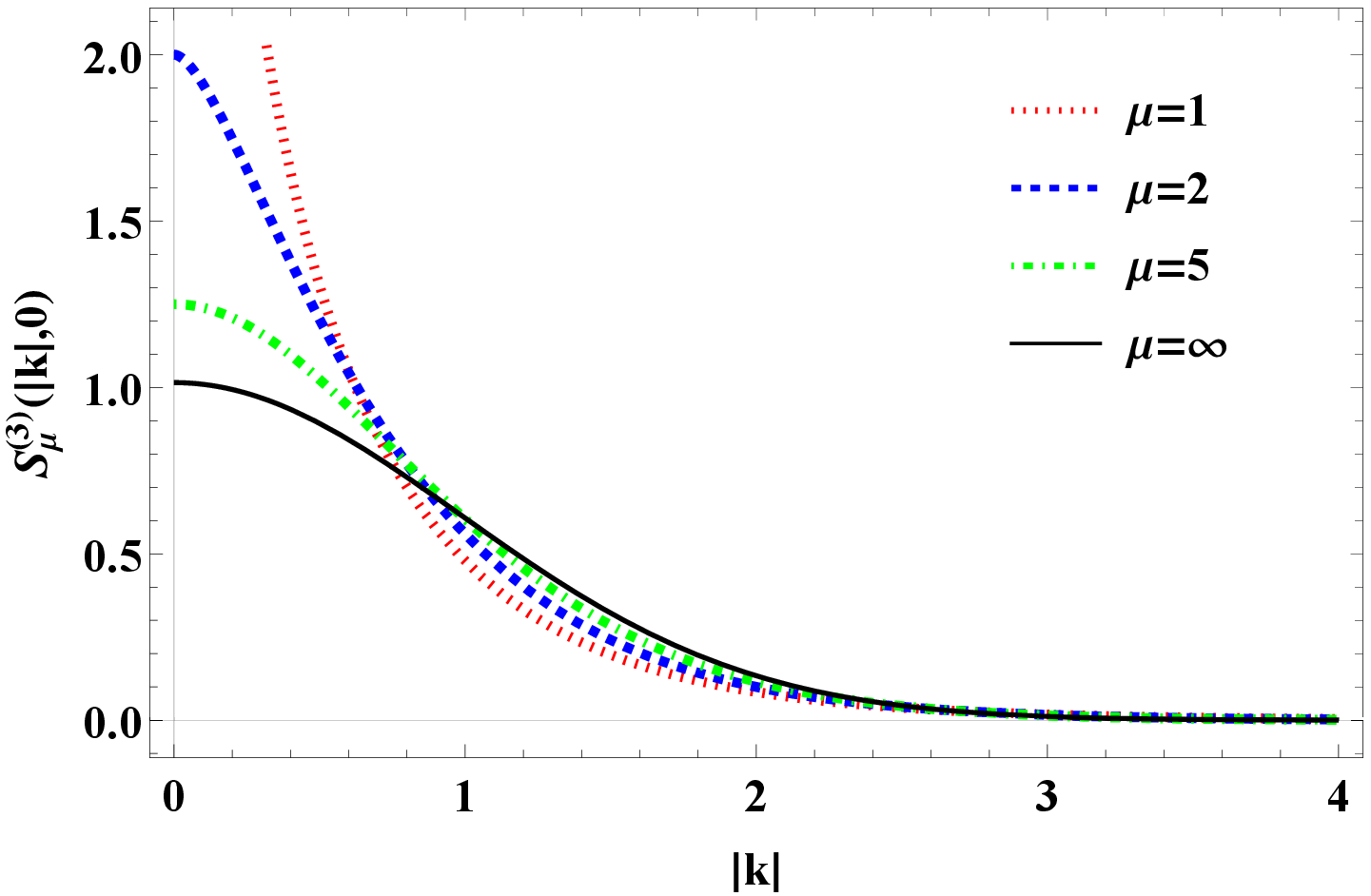}%
	}	\hspace*{0.1cm}
	\subfloat[\label{fig_1f}]{
		\includegraphics[width=.48\linewidth]{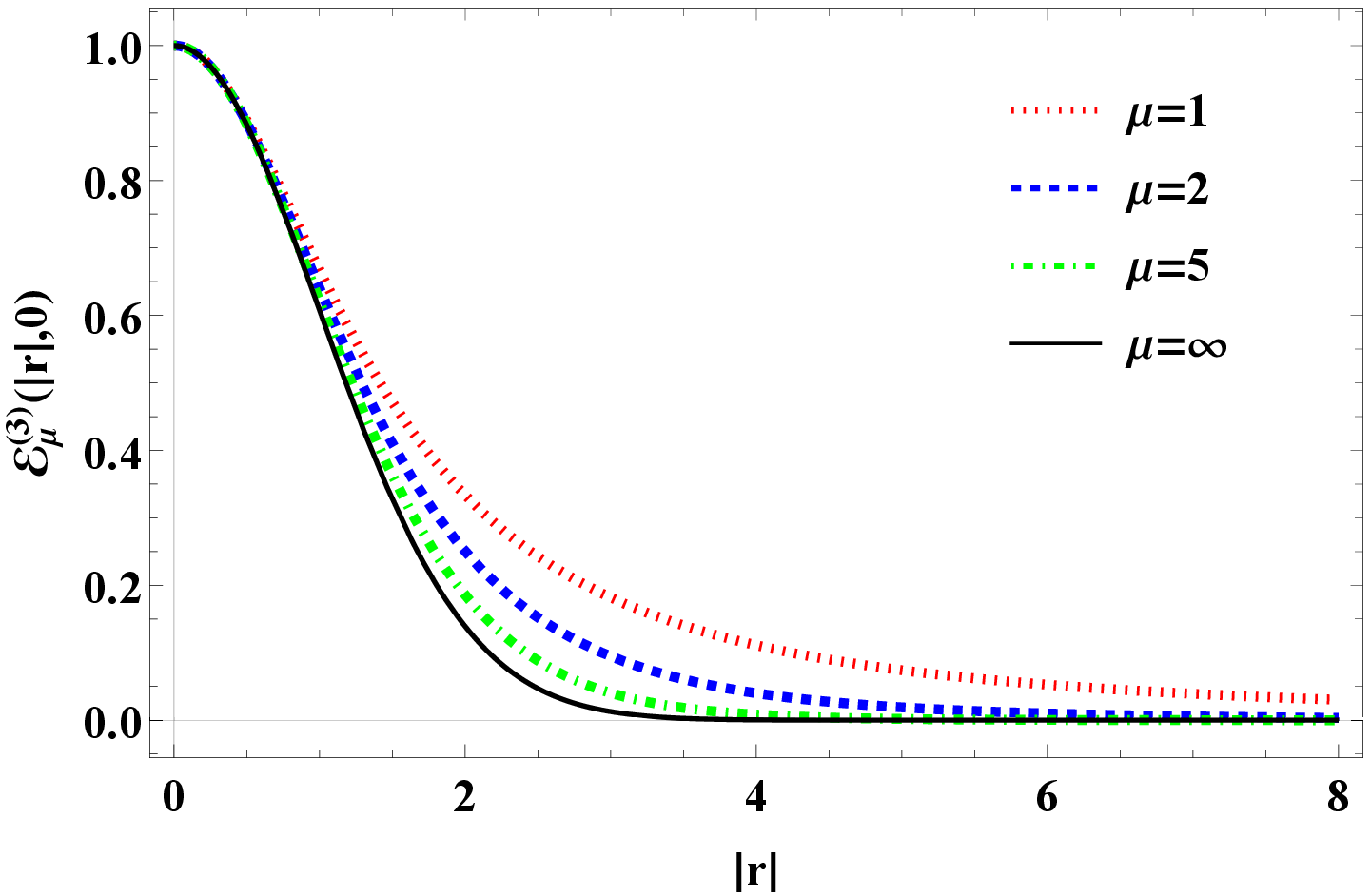}%
	}
	\caption{Spectra $S_{\mu}^{(3)}(|\mathbf{k}|,0)$ and the correlation functions $\mathcal{E}_{\mu}^{(3)}(|\mathbf{r}|,0)$ for several values of the parameter $\mu$ and $\tau_c=1,\lambda_c=1$. The black lines correspond to the Gaussian limit.}
\end{figure}

The correlation $\mathcal{E}_{\mu}^{(3)}(\mathbf{x},t)$ depicts algebraically decaying space correlation and it is able to encompass both the short-ranged ($\mu\to \infty$) and the long-ranged ($\mu\to 1$) limits. In Figs. \ref{fig_1e}-\ref{fig_1f} we plot $\mathcal{E}_{\mu}^{(3)}(|\mathbf{r}|,0)$ and $S_{\mu}^{(3)}(|\mathbf{k}|,0)$ at different $\mu$ values, setting $\lambda_c=1$.

\begin{equation}
	\label{eq_4}
	S_{\mu}^{(3)}(\mathbf{k},\omega) \sim \left| \mathbf{k}\lambda_c \right| ^{\mu -1} K_{\mu -1}\left(\sqrt{2\mu } \left| \mathbf{k}\lambda_c \right| \right)e^{-\frac{\omega^2\tau_c^2}{2}}\longrightarrow \mathcal{E}_{\mu}^{(3)}(\mathbf{x},t)= e^{-\frac{t^2}{2\tau_c^2}}(1+\frac{|\mathbf{x}|^2}{2\lambda_c^2\mu})^{-\mu}.\end{equation}

At this end, we reiterate that $\mathcal{E}^{(1)}_\nu,\mathcal{E}^{(2)}_\mu,\mathcal{E}^{(3)}_\mu$ are designed to describe long-range time, oscillating space, respectively, long-range space correlations. Supplementary, their functional forms were chosen such that, in specific limits, the following equality holds $\mathcal{E}^{(1)}_\infty(\mathbf{x},t)=\mathcal{E}^{(2)}_0(\mathbf{x},t)=\mathcal{E}^{(3)}_\infty(\mathbf{x},t)$.

\subsection{Kubo numbers}
\label{section_2.2}

It is customary in turbulence studies to define a so-called Kubo number $K_\star$ \cite{kubo,PCastiglione_2000} as the ratio between a characteristic time-scale of turbulence $T$ and the time-of-flight $\tau_{fl}$,  $K_\star=T/\tau_{fl}$. The latter is understood as the specific time in which the space-correlation is lost and can be approximated as the ratio between a characteristic space-scale of turbulence $L$ and the average amplitude of the velocity field $V$, $\tau_{fl} = L/V$. The Kubo number $K_\star=TV/L$ is also a measure of the turbulence amplitude, via $V$. From a physical point of view, $K_\star$ indicates the ability of individual particles to explore the spatial structure of the velocity field before the latter changes beyond recognition, i.e. becomes decorrelated. 

With respect to $K_\star$, three distinct regimes with different behaviours of the asymptotic diffusion $D^\infty$ can be defined: the quasilinear regime (weak turbulence or high frequency) when $K_\star\ll 1$ and $D^\infty \sim K_\star^2$, the strong regime (strong turbulence, intermediate frequencies) when $K_\star\sim 1$ and $D^\infty \sim K_\star^1$ and the percolative regime (strong turbulence, low-frequency) when $K_\star\gg 1$ and $D^\infty\sim K_\star^\gamma$. For the latter case Isichenko derived his famous result of $\gamma = 7/10$ whose universality is under scrutiny in this paper.

Before searching for the relation between Kubo numbers and diffusion, a question is to be asked: what is the correct way to define $K_\star$, or equivalently, what are the \emph{characteristic time-space scales} of turbulence, $T,L$? One possible answer is to consider that these scales are the Taylor differential scales $(T_D,L_D)$ \cite{monin2013statistical}:
\begin{align}
	\label{eq_5}
	T\equiv T_D(\nu)  &= \left(-\frac{\mathcal{E}(\mathbf{0},0)}{\partial_{t,t}\mathcal{E}(\mathbf{0},0)}\right)^{1/2},\\
	L\equiv L_D(\mu) &= \left(-\frac{\mathcal{E}(\mathbf{0},0)}{\partial_{x,x}\mathcal{E}(\mathbf{0},0)}\right)^{1/2}.
\end{align}

In fact, the parametric dependencies of $\mathcal{E}^{(1-3)}$ were chosen in such a way that all have the same differential scales  $L_D = \lambda_c, T_D=\tau_c$ and the same Eulerian velocity field amplitude $V= \sqrt{\langle\mathbf{v}^2(\mathbf{x},t)\rangle}=\sqrt{-\nabla^2\mathcal{E}(\mathbf{x},t)}=\lambda_c^{-1}$, consequently, the same \emph{differential Kubo number $K_D=VT_D/L_D=\tau_c/\lambda_c^2$}. 

On the other hand, the differential $L_D, T_D$, as the name suggests, describe only the local tendency of correlation and may not say anything relevant about the global features which might play a decisive role in strong transport. Moreover, all functions $\mathcal{E}^{(1-3)}$, when departing from the Gaussian limit, lose their monoscale character. For that, we define here the integral time-space scales $(T_I,L_I)$ \cite{monin2013statistical} as global measures of the correlation:

\begin{align}
\label{eq_6}
	L\equiv L_I(\mu) &=\int_0^\infty \frac{\mathcal{E}(r,0)}{\mathcal{E}(\mathbf{0},0)}~dr  \\
	T\equiv T_I(\nu) &=\int_0^\infty \frac{\mathcal{E}(\mathbf{0},t)}{\mathcal{E}(\mathbf{0},0)}~dt. 
\end{align}
The correlations $\mathcal{E}^{(1-3)}$ allow us to evaluate these quantities $(T_I,L_I)$ analytically and see that they coincide with $\sqrt{\pi/2} (T_D,L_D)$ in specific asymptotic limits:
\begin{align*}
	T_I(\nu) &= \tau_c\sqrt{\frac{\pi\nu }{2}}\frac{ \Gamma \left(\nu -\frac{1}{2}\right)}{\Gamma (\nu )}\longrightarrow[\nu\to\infty]{} \sqrt{\frac{\pi}{2}} \tau_c, \hspace{0.5cm} &\mathrm{for}\hspace{0.5cm}  \mathcal{E}_{\nu}^{(1)}\\
	L_I(\mu) &= \lambda_c\sqrt{\frac{(\mu +2)}{\pi\mu^2}} \frac{\Gamma \left(\frac{\mu +1}{2}\right)}{\Gamma \left(\frac{\mu }{2}\right)}\longrightarrow[\mu\to 0]{}\sqrt{\frac{\pi}{2}} \lambda_c, \hspace{0.5cm} &\mathrm{for}\hspace{0.5cm} \mathcal{E}_{\mu}^{(2)}\\
	L_I(\mu) &= \lambda_c\sqrt{\frac{\pi \mu}{2}}\frac{ \Gamma \left(\mu -\frac{1}{2}\right)}{\Gamma (\mu )}\longrightarrow[\mu\to \infty]{} \sqrt{\frac{\pi}{2}}\lambda_c, \hspace{0.5cm} &\mathrm{for}\hspace{0.5cm}  \mathcal{E}_{\mu}^{(3)}.
\end{align*}

Thus, we can define, as an alternative to differential $K_D$, the integral Kubo number $K_I$

$$K_I = \frac{V T_I}{L_I}.$$

\subsection{Direct numerical simulations}
\label{section_2.3}
 
Regarding the specific numerical approach on computing the diffusion coefficients, we employ here the so-called direct numerical simulation method (DNS) \cite{Palade2021,Palade_2021,Vlad_2021}. DNS attempts to mimic the real trajectories $\{\mathbf{x}(t)\}$ at the numerical level. In practice, a finite ensemble (dimension $N_p$) of random fields $\{\phi(\mathbf{x},t)\}$ is generated. For each such realization, the associated trajectory $\mathbf{x}(t)$ that obeys eq. \eqref{eq_1} is computed. The technical method of generating a Gaussian Random Field with given spectrum $S(\mathbf{k},\omega)$ is via a Fourier decomposition:

\begin{align}
	\label{eq_7}
	\phi(\mathbf{x},t) = \sqrt{\frac{2}{N_c}}\sum_j^{N_c} sin(\mathbf{k}_j\mathbf{x}-\omega_j t+\zeta_j),
\end{align}
where $\zeta_j$ are random phases $\in (0,2\pi]$, while the wave-vectors $\mathbf{k}_j$ and the frequencies $\omega_j$ are independent random variables that have as PDF the turbulence spectrum $S(\mathbf{k},\omega)$. The resulting field will be zero-averaged $\langle\phi(\mathbf{x},t)\rangle = 0$ and will have the correct correlation function $\langle\phi(\mathbf{x},t)\phi(\mathbf{x}^\prime,t^\prime)\rangle=\mathcal{E}\left(\mathbf{x}-\mathbf{x}^\prime,t-t^\prime\right)$ associated to the spectrum $S(\mathbf{k},\omega)$ by construction. Its Gaussianity is guaranteed by the Central Limit Theorem for large numbers of waves $N_c\to \infty$. For more details see \cite{Palade2021}.

In practice, we use high-fidelity simulations with unprecedented numerical resolution. The dimension of the statistical ensemble is set to $N_p = 10^7$ trajectories and $N_c=10^3$ partial waves are used. In this way we ensure proper Gaussianity of the turbulent field, a good convergence to the homogeneous correlation, the lack of spurious correlations or periodicity, as well as a good convergence of the Lagrangian ensemble of trajectories over large times. The latter are computed solving the eq. \eqref{eq_1} with a 4th order Runge-Kutta method and a fixed time-step of $\delta t\approx 0.05$. This is enough to achieve the stability of trajectories and their Hamiltonian character (when present) over long periods of time, which usually are set to $t_{max}\sim 5-10\tau_c$. Note that, in order to simulate different values of Kubo numbers (either differential $K_D$ or integral $K_I$), we keep $\lambda_c=1\implies V=1$ in all simulations, but vary the $\tau_c$ parameter. Using dedicated programming procedures, typical simulations on several processor servers require in terms of CPU time: $t_{CPU}\sim 1\mathrm{min}$ for $\tau_c\sim 1$ and up-to $t_{CPU}\sim 10^2 \mathrm{min}$ for $\tau_c\sim 10^3$.

\section{Results}
\label{section_3}

\subsection{Generic features of diffusion processes}
\label{section_3.1}

The diffusion coefficients $D_x(t)$ are intimately related to the mean square displacement $\sigma_x(t)$ and the Lagrangian velocity autocorrelation $\mathcal{L}_{x}(t)=\langle v_x(0)v_x(t)\rangle$ through iterative time-differentiation $\mathcal{L}_x(t) = d_tD_x(t) =1/2 d_{t,t}\sigma_x(t)$. Consequently, algebraic dependencies are expected as long-time behaviour $\lim\limits_{t\to\infty}\sigma_x(t)\sim t^{\alpha},D_x(t)\sim t^{\alpha-1},\mathcal{L}_x(t)\sim t^{\alpha-2}$. The turbulent transport in 2D incompressible fields lies between two limits: \emph{frozen} and \emph{quasilinear}.

\begin{figure}
	\subfloat[\label{fig_0}]{
		\includegraphics[width=.93\linewidth]{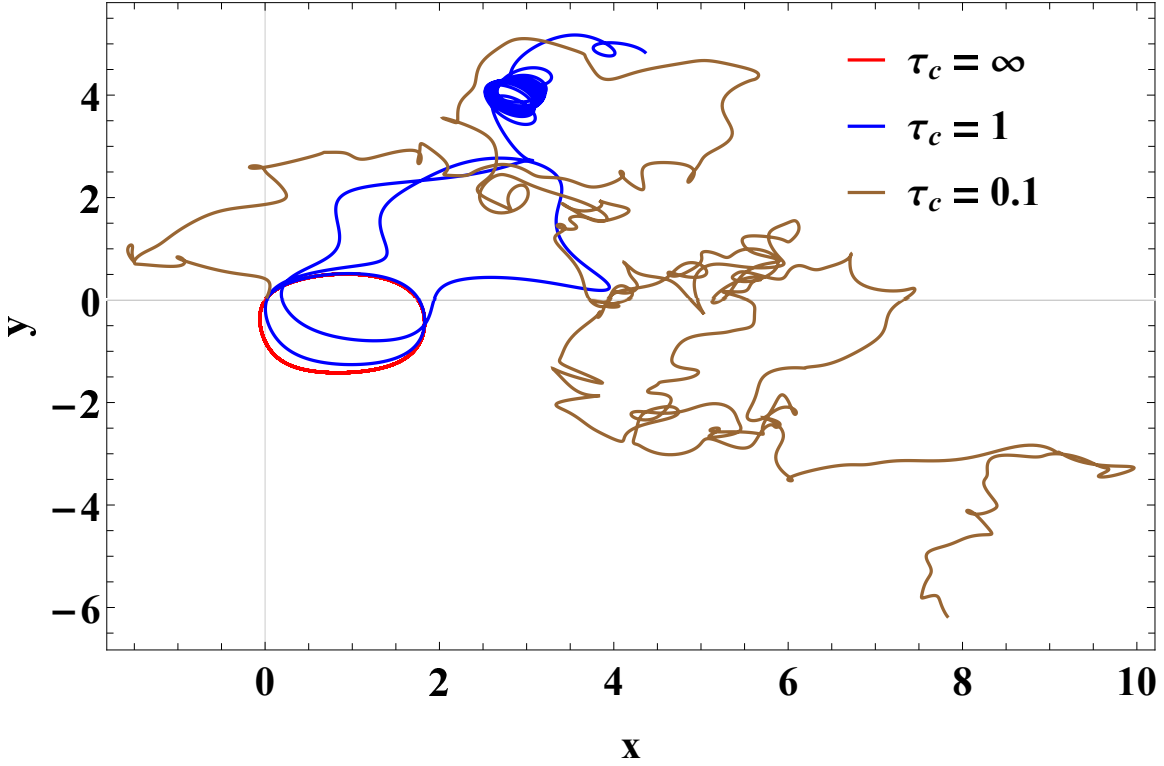}}
	\caption{Typical trajectories $\mathbf{x}(t)$ in the same field $\phi(\mathbf{x},t/\tau_c)$ with $\tau_c = \{\infty, 1, 0.1\}$ (red, blue, brown). }
\end{figure}

\begin{figure}
	\subfloat[\label{fig_2a}]{
	\includegraphics[width=.48\linewidth]{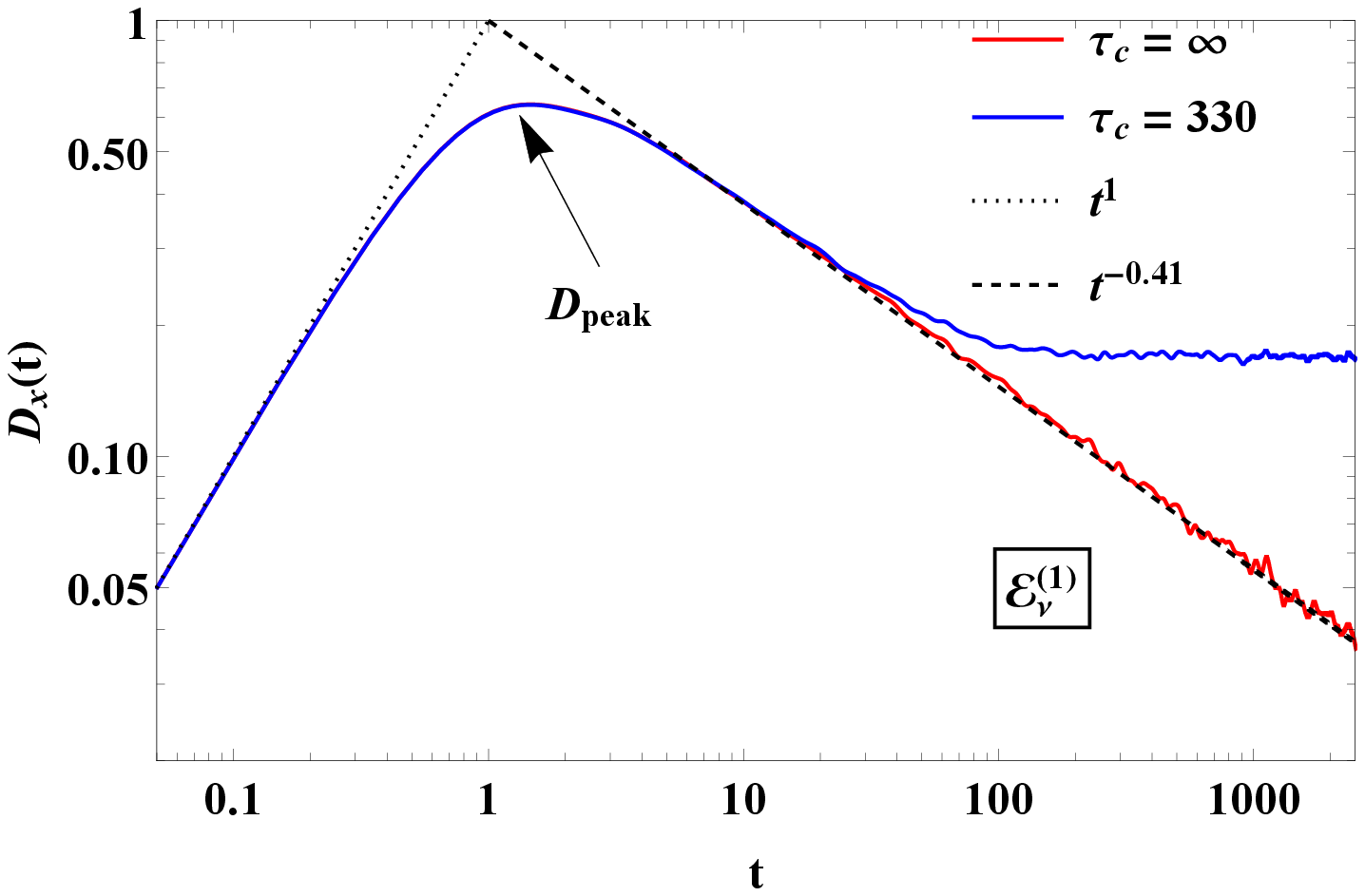}%
}\hspace*{0.1cm}
\subfloat[\label{fig_2b}]{
	\includegraphics[width=.48\linewidth]{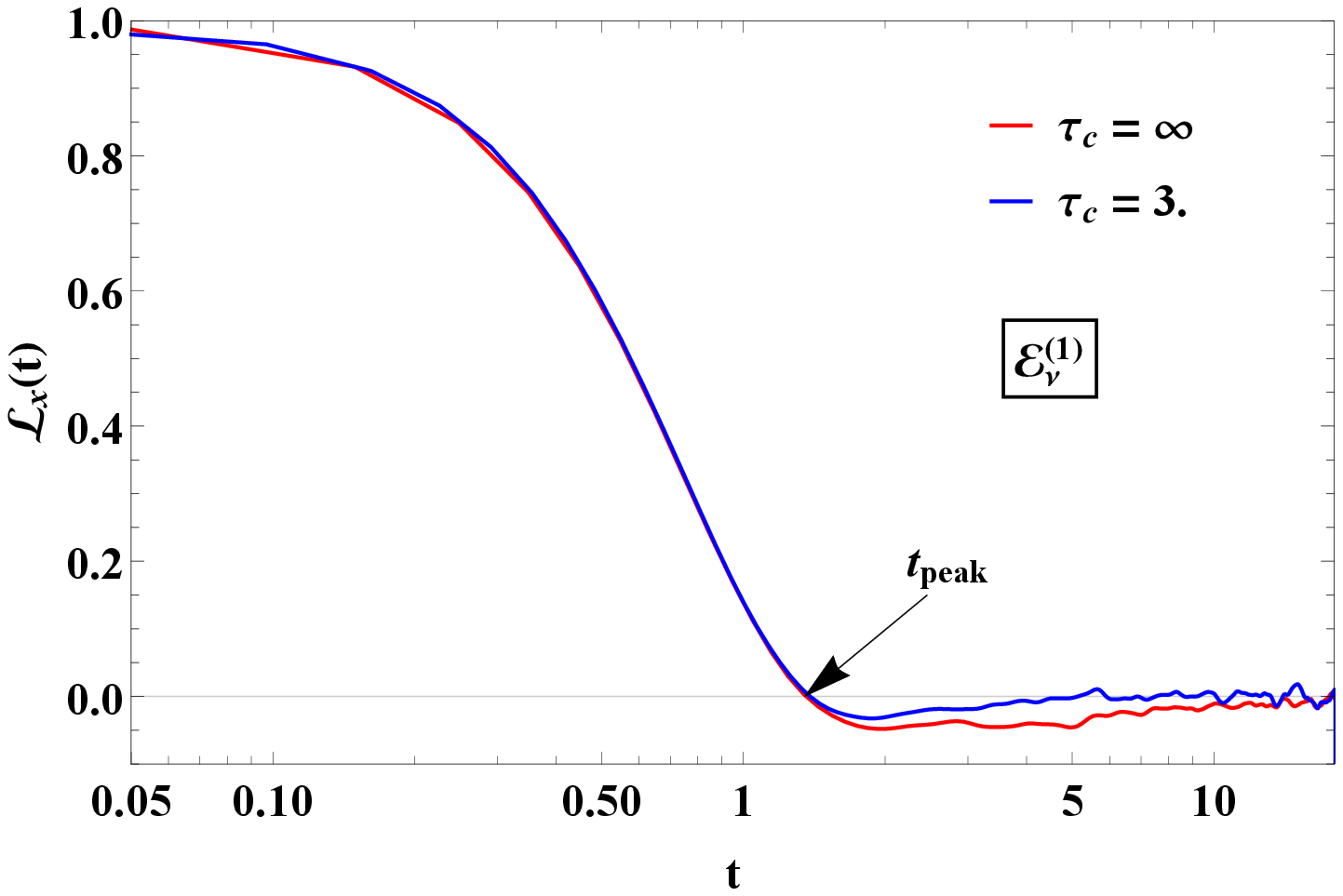}%
}
	\caption{Running diffusion coefficient $D_x(t)$ (left, a)) and Lagrangian correlation $\mathcal{L}_{x}(t)$ (right, b)) for the correlation $\mathcal{E}_{\nu}^{(1)}(\mathbf{x},t)$ at $\nu=5, \lambda_c=1$ and two different values of $\tau_c = 3, 330$ (blue line) and $\tau_c =\infty$ (red line).}
\end{figure}

The \emph{frozen turbulence} limit corresponds to stationary fields, i.e. $\mathbf{v}(\mathbf{x},t)\equiv \mathbf{v}(\mathbf{x})$, $\phi(\mathbf{x},t)\equiv \phi(\mathbf{x})$, when $\tau_c\to \infty$. Due to the Hamiltonian nature of the equations of motion \eqref{eq_1}, it follows that the trajectories are closed. This can be seen in Fig. \eqref{fig_0}, where a generic stationary trajectory is drawn in red. At small times, the Lagrangian velocities experienced by particles are roughly $\mathbf{v}\left(\mathbf{x}(t),t\right)\approx \mathbf{v}\left(\mathbf{x}(0),0\right)$ which enables us to show that $\sigma_x(t)\approx V^2 t^2, D_x(t)\approx V^2t, \mathcal{L}_x(t)\approx V^2$. At long times, since all particles are trapped in equipotential field-lines, their asymptotic displacement $\lim\limits_{t\to\infty}\sigma_x(t)$ is finite, thus, the diffusion coefficient $\lim\limits_{t\to\infty}D_x(t)$ and the Lagrangian correlation $\mathcal{L}_x(t)$ decay to zero. At intermediate times, at the so called $t_{peak}\approx \tau_{fl}$, the Lagrangian correlation has a first zero which corresponds to a maximum in diffusion $D_{peak}$. All these features can be observed in Figs.
\eqref{fig_2a},\eqref{fig_2b} for the red lines corresponding to frozen running diffusion $D_x(t)$ (left) and Lagrangian correlation $\mathcal{L}_{x}(t)$ (right) in the case of $\mathcal{E}^{(1)}_\nu$. The numerical simulations confirm the anomalous algebraic dependence $D_x(t)\sim t^{\alpha-1}$ with the exponent $\alpha\approx 0.6$.

The existence of decorrelation mechanisms (through a finite $\tau_c$) induces a continuous change on the landscape of the stream-function $\phi(\mathbf{x},t)$ which is equivalent, at the level of individual trajectories, with long jumps between equipotential lines (Fig. \eqref{fig_0}, blue line). This implies that the particles are no longer trapped, and their spreading is continuously increasing. The Lagrangian correlation $\mathcal{L}_{x}(t)$ decays strongly to $0$ which makes the process diffusive, i.e. $\alpha\to 1$ and $D_x^\infty $ finite. These features can be seen in 
\eqref{fig_2a},\eqref{fig_2b} for the blue lines which correspond to finite $K_\star$ regimes.

The \emph{quasilinear} limit is characterized by a small Kubo number $K_\star\ll 1$ (or decorrelation time $\tau_c\ll t_{peak}$). The particles are driven by fields that change so rapidly in time that their motion is almost unaffected by the space structure of the fields. The trajectories exhibit coloured-noise driven features (Fig. \eqref{fig_0} brown line) and the diffusion coefficient can be estimated as the product between field strength $V^2$ and the integral time scale $T_I$: 

$$D_x^\infty=\int_0^\infty \mathcal{L}_x(t)~dt = \int_0^\infty \langle v_x(\mathbf{x}(0),0)v_x(\mathbf{x}(t),t)\rangle \approx \int_0^\infty \langle v_x(\mathbf{0},0)v_x(\mathbf{0},t)\rangle = V^2 T_I.$$

The diffusion profiles shown in Figs. \eqref{fig_2a} are generic for all correlation functions $\mathcal{E}^{(1-3)}$, both in the frozen $\tau_c\to\infty$ or decorrelated $\tau_c = $ finite regimes. 

We underline that the results from all figures shown in the present work are scaled as it follows: $t\equiv t V/L_D, \mathcal{L}_x\equiv \mathcal{L}_x/V^2, D_x\equiv D_x/(L_DV), L_I\equiv L_I/L_D, T_I\equiv T_I/T_D$. Nonetheless, throughout the text, all formulae contain the correct, non-scaled dependencies.

\subsection{Frozen turbulence}
\label{section_3.2}

\begin{figure}
	\subfloat[\label{fig_3a}]{
		\includegraphics[width=.48\linewidth]{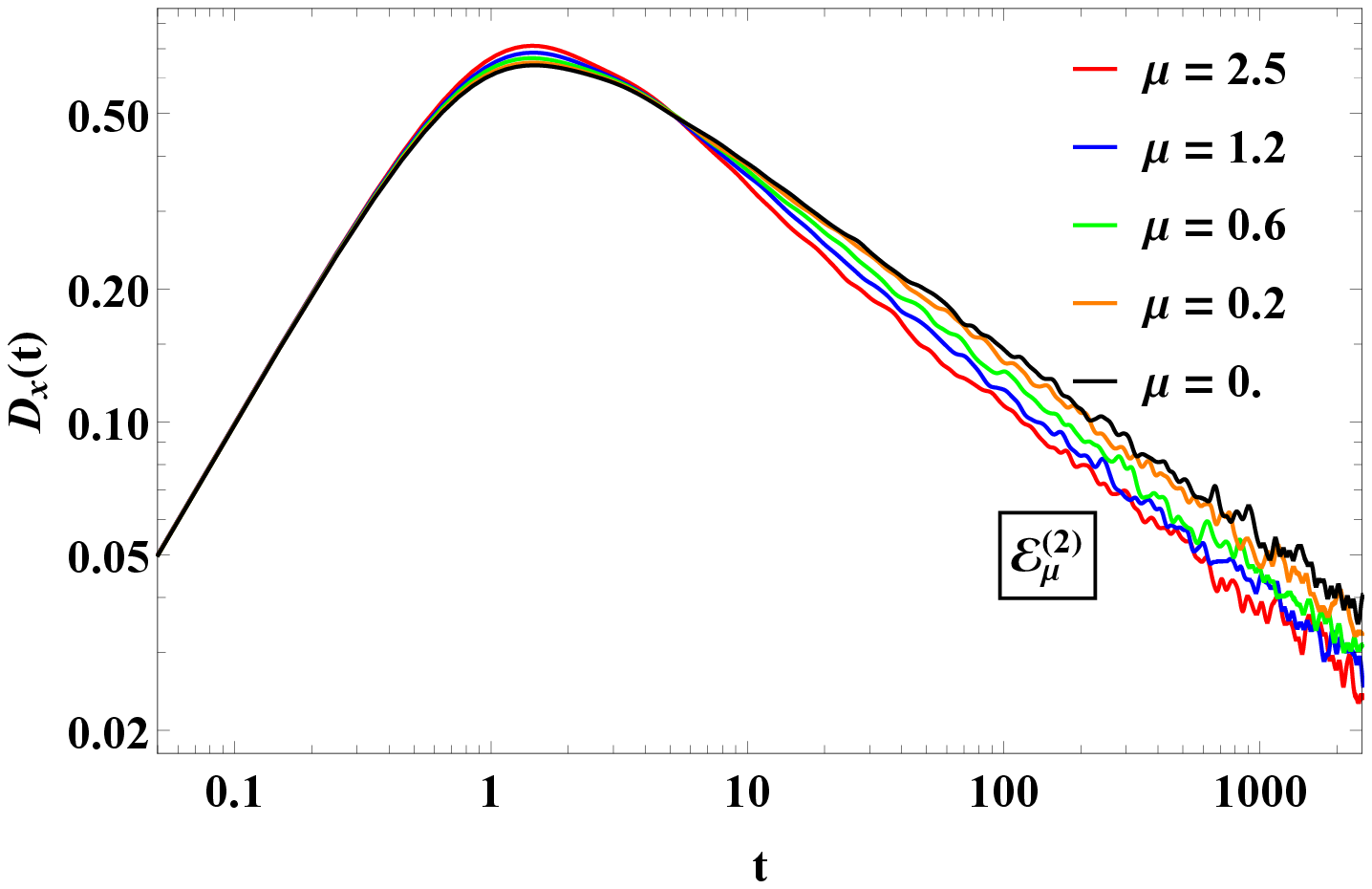}%
	}\hspace*{0.1cm}
	\subfloat[\label{fig_3b}]{
		\includegraphics[width=.48\linewidth]{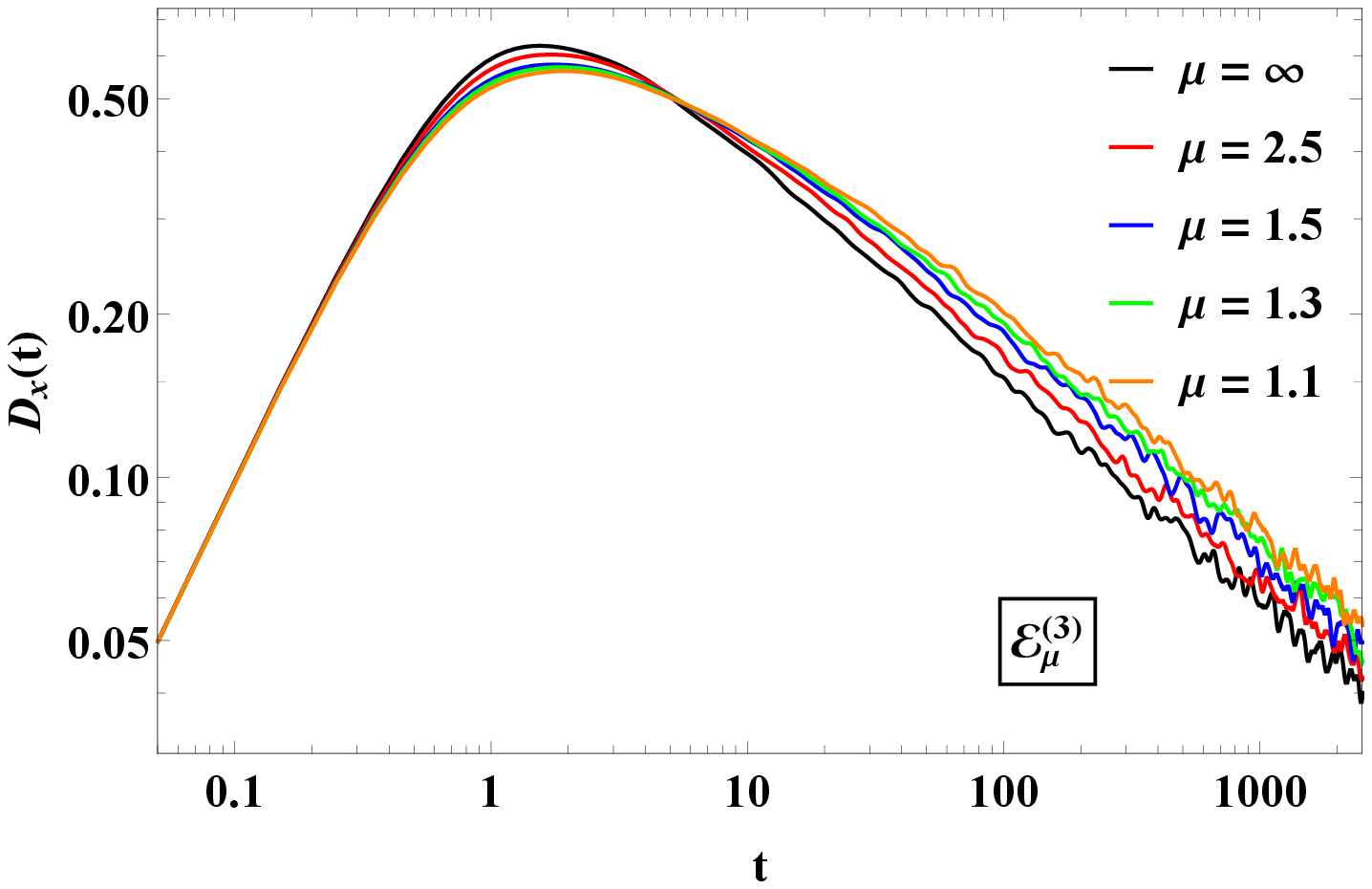}%
	}
\caption{Running diffusion coefficients $D_x(t)$ in the case of frozen turbulence $\lambda_c=1, \tau_c\to\infty$ for the correlations $\mathcal{E}_{\mu}^{(2)}(\mathbf{r},0)$ (left) and $\mathcal{E}_{\mu}^{(3)}(\mathbf{r},0)$ (right) at different $\mu$ values. Note that the black lines ($\mu = 0$ (left) or $\mu=\infty$ (right)) represents the Gaussian limit.}
\end{figure}

\begin{figure}
	\subfloat[\label{fig_4a}]{
		\includegraphics[width=.48\linewidth]{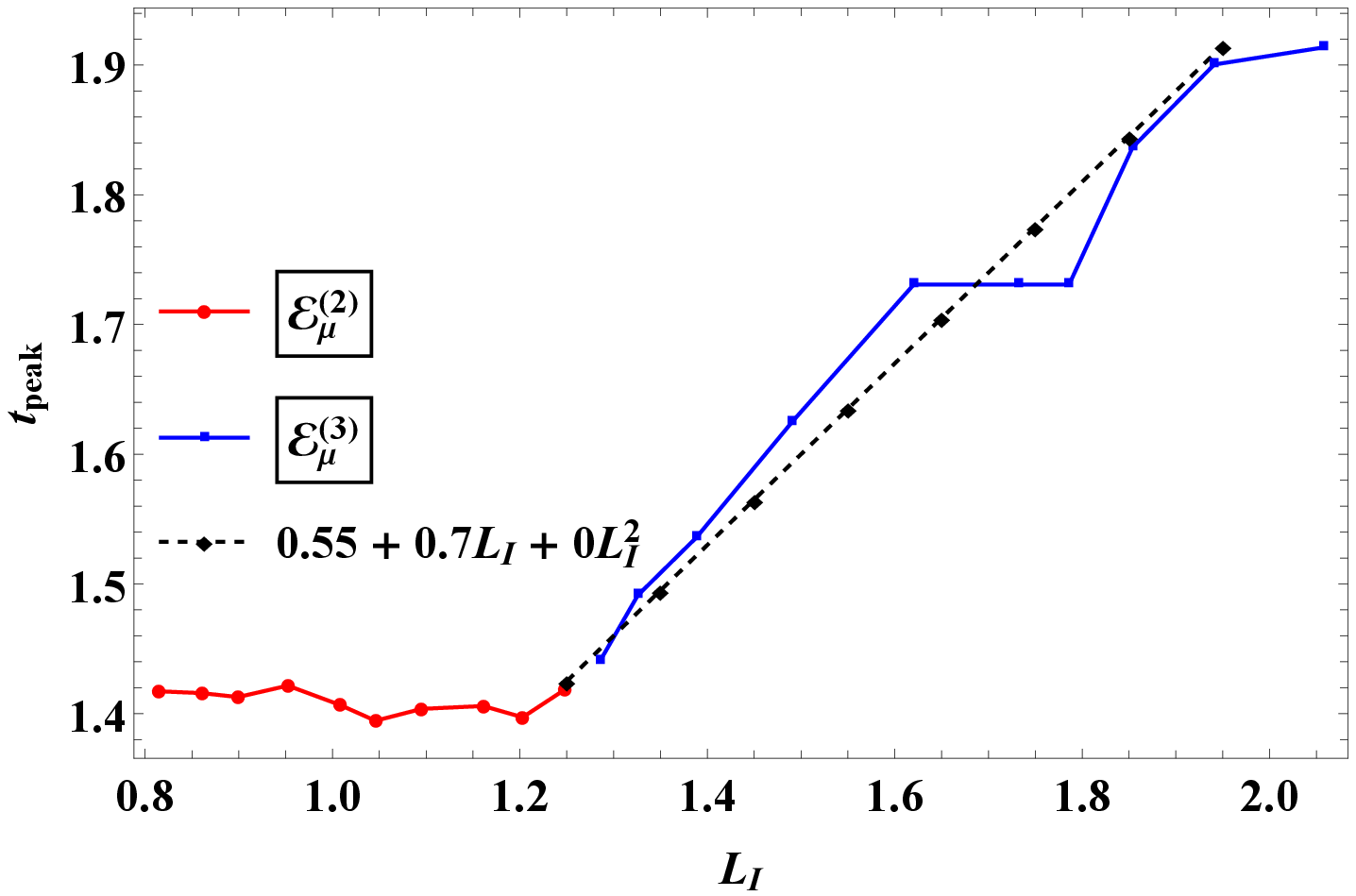}%
	}\hspace*{0.1cm}
	\subfloat[\label{fig_4b}]{
		\includegraphics[width=.48\linewidth]{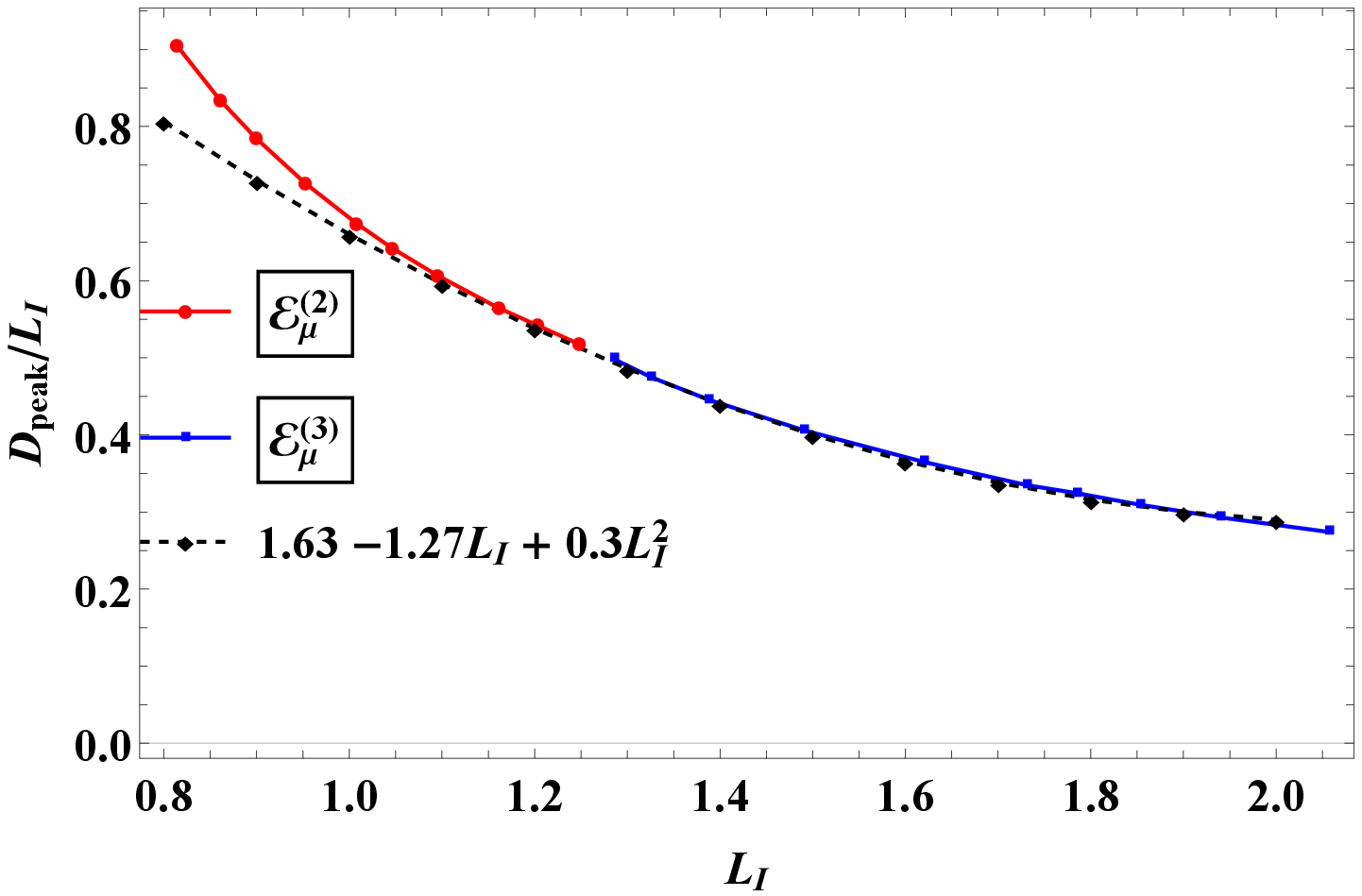}%
	}\\
	\subfloat[\label{fig_4c}]{
		\includegraphics[width=.48\linewidth]{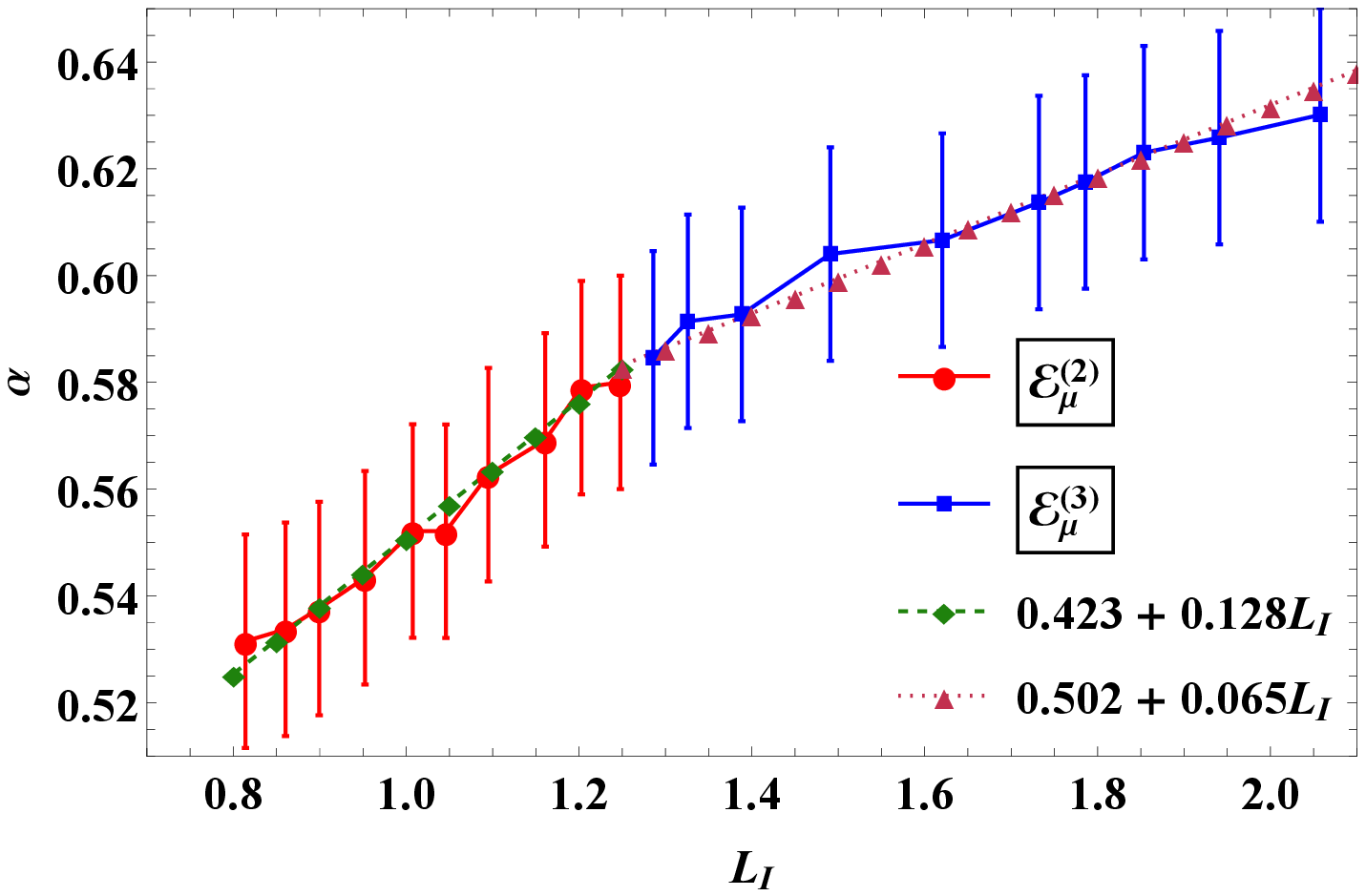}%
	}\hspace*{0.1cm}
	\subfloat[\label{fig_4d}]{
	\includegraphics[width=.48\linewidth]{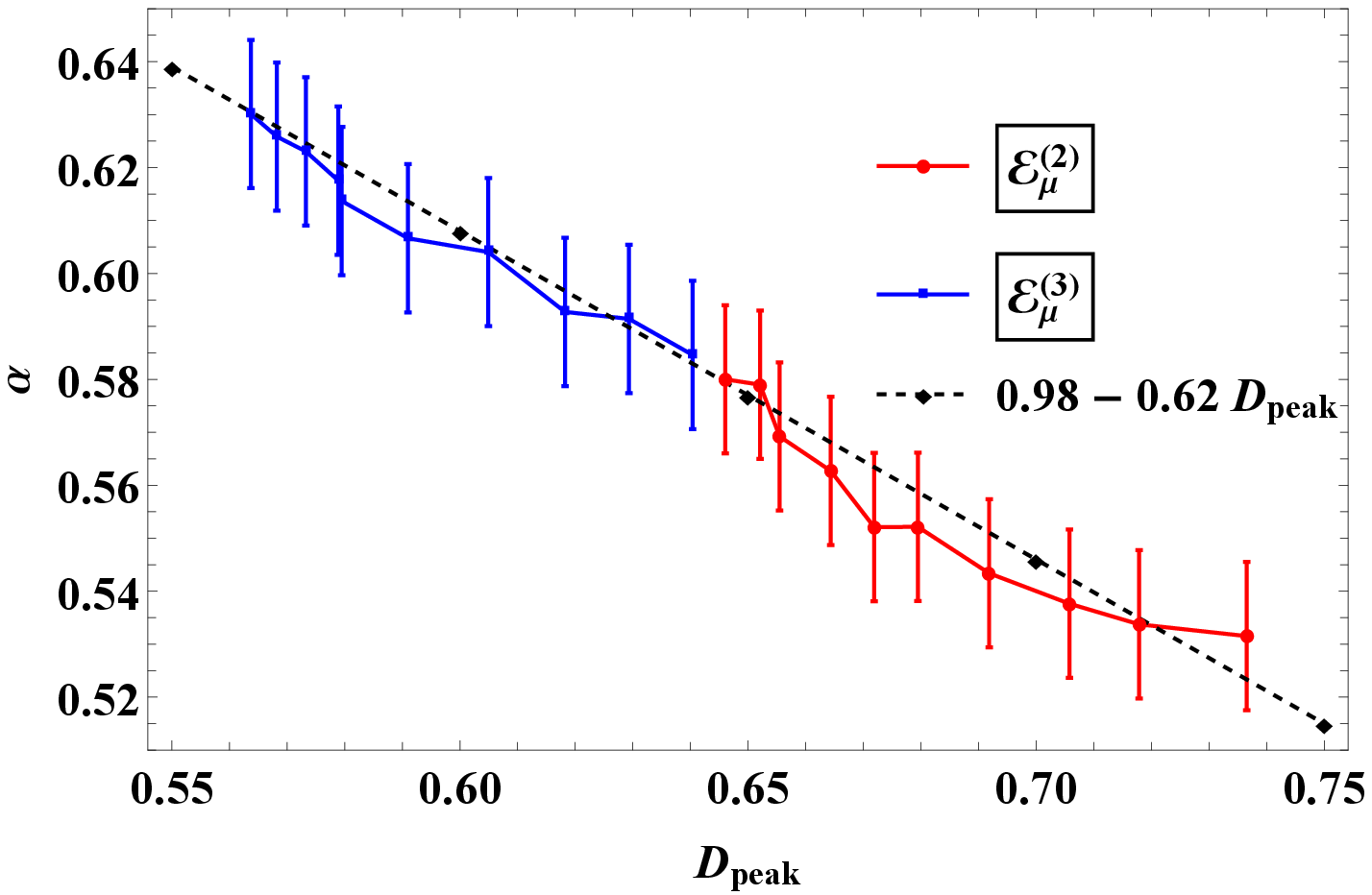}%
}
	\caption{$\alpha$ anomalous exponent for the correlation classes $\mathcal{E}^{(2)}$ (red) and $\mathcal{E}^{(3)}$ (blue) vs their integral correlation length. Note the error bars are estimated from the amplitude of numerical fluctuations present in the tail of the diffusion coefficients.}
\end{figure}

We consider now, in detail, the diffusion process in the case of frozen turbulence, which is an ideal limit of very slowly varying velocity fields $\tau_c\to \infty$. The behaviour apparent in Fig. \eqref{fig_2a} is qualitatively relevant for all correlations and can be observed in Figs. \eqref{fig_3a}-\eqref{fig_3b} where $D_x(t)$ is represented for $\mathcal{E}_{\mu}^{(2)}$, respectively $\mathcal{E}_{\mu}^{(3)}$ at different values of the $\mu$ parameter. At small times $t\ll t_{peak}$ all profiles follow the analytical behaviour $D_x(t)\approx V^2 t^1$ discussed in the previous section \eqref{section_3.1}. Maximal values $D_{peak}$ are present at $t_{peak}$, whereupon, diffusion decays algebraically $D_x(t)\sim t^{\alpha-1}$. Thus, these three quantities $t_{peak},D_{peak},\alpha$ and their relation with the functional structure of turbulence are of interest for the frozen regime. Since all correlations have the same differential scales $L_D=\lambda_c$, we discern between them with the aid of the integral scale length $L_I$. We represent the dependencies $t_{peak}(L_I)$, $D_{peak}(L_I)$, $\alpha(L_I)$ in Figs. \eqref{fig_4a}-\eqref{fig_4c}.

The time at which the maximal diffusion is reached, $t_{peak}(L_I) \sim \sqrt{2}\lambda_c/V\approx L_I/V$, can be viewed as the \emph{real time-of-flight} and it turns out Fig. \eqref{fig_4a} to be consistently larger than the gross differential estimation $\tau_{fl}=L_D/V = \lambda_c^2\to 1$. Moreover, it appears that the oscillating nature of $\mathcal{E}^{(2)}$ has virtually no effect (apart from numerical fluctuations) on $t_{peak}$. In contrast, the long-range character of $\mathcal{E}^{(3)}$ leads to an approximate linear increase:

$$t_{peak}\approx \frac{L_D}{V}\left(0.55+0.7\left(\frac{L_I}{L_D}\right)\right).$$

The maximal value of the diffusion, $D_{peak}$, is influenced both by oscillations and long-range features of the space correlations. The dependencies represented in Fig. \eqref{fig_4b} resemble a relatively flat quadratic function, thus we estimate: 

$$D_{peak}\approx VL_I\left(1.63-1.27 \left(\frac{L_I}{L_D}\right)+0.3\left(\frac{L_I}{L_D}\right)^2\right).$$

Finally, we extract from the numerical data the percolative exponent $\alpha$ which dictates the long-time behaviour of diffusion $D_x(t)\approx V^{\alpha}L_D^{2-\alpha}t^{\alpha-1}$ and find that it has almost linear dependencies with the integral scale $L_I$ and the maximal diffusion $D_{peak}$, but with different slopes for $\mathcal{E}^{(2)}$ and $\mathcal{E}^{(3)}$. We approximate $\alpha \approx 0.423+0.128L_I/L_D$ for oscillating $\mathcal{E}^{(2)}$ and $\alpha \approx 0.502+0.065L_I/L_D$ for long-range $\mathcal{E}^{(3)}$, while $\alpha\approx 0.98-0.62D_{peak}/(VL_D)$. Note that error bars are represented in Figs. \eqref{fig_4c},\eqref{fig_4d} as an estimation of the numerical error induced by the finite statistical ensemble of trajectories and the fitting procedure.

\subsection{Time dependent turbulence}
\label{section_3.3}

\begin{figure}
	\subfloat[\label{fig_6a}]{
		\includegraphics[width=.33\linewidth]{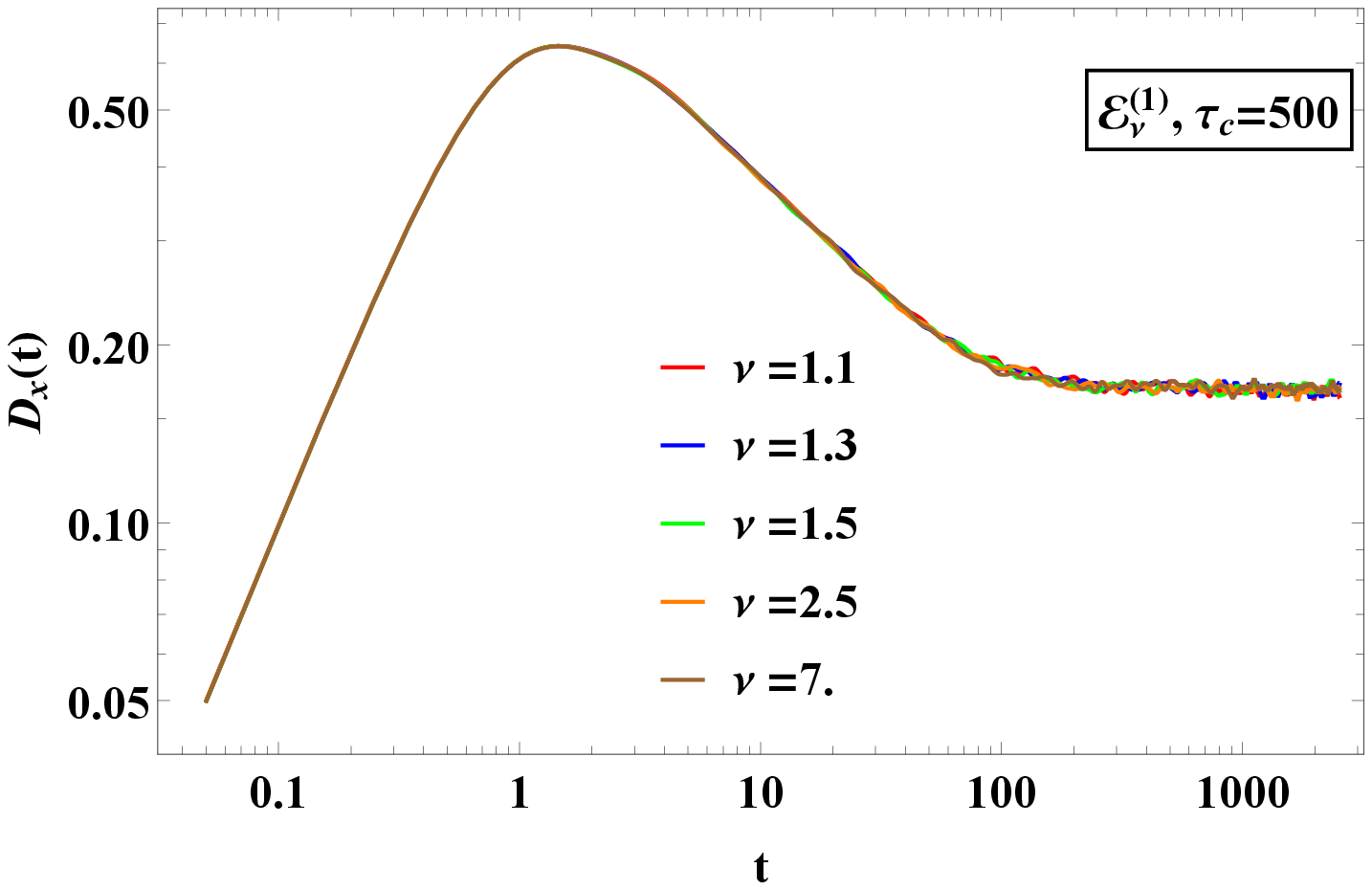}%
	}
	\subfloat[\label{fig_6b}]{
		\includegraphics[width=.33\linewidth]{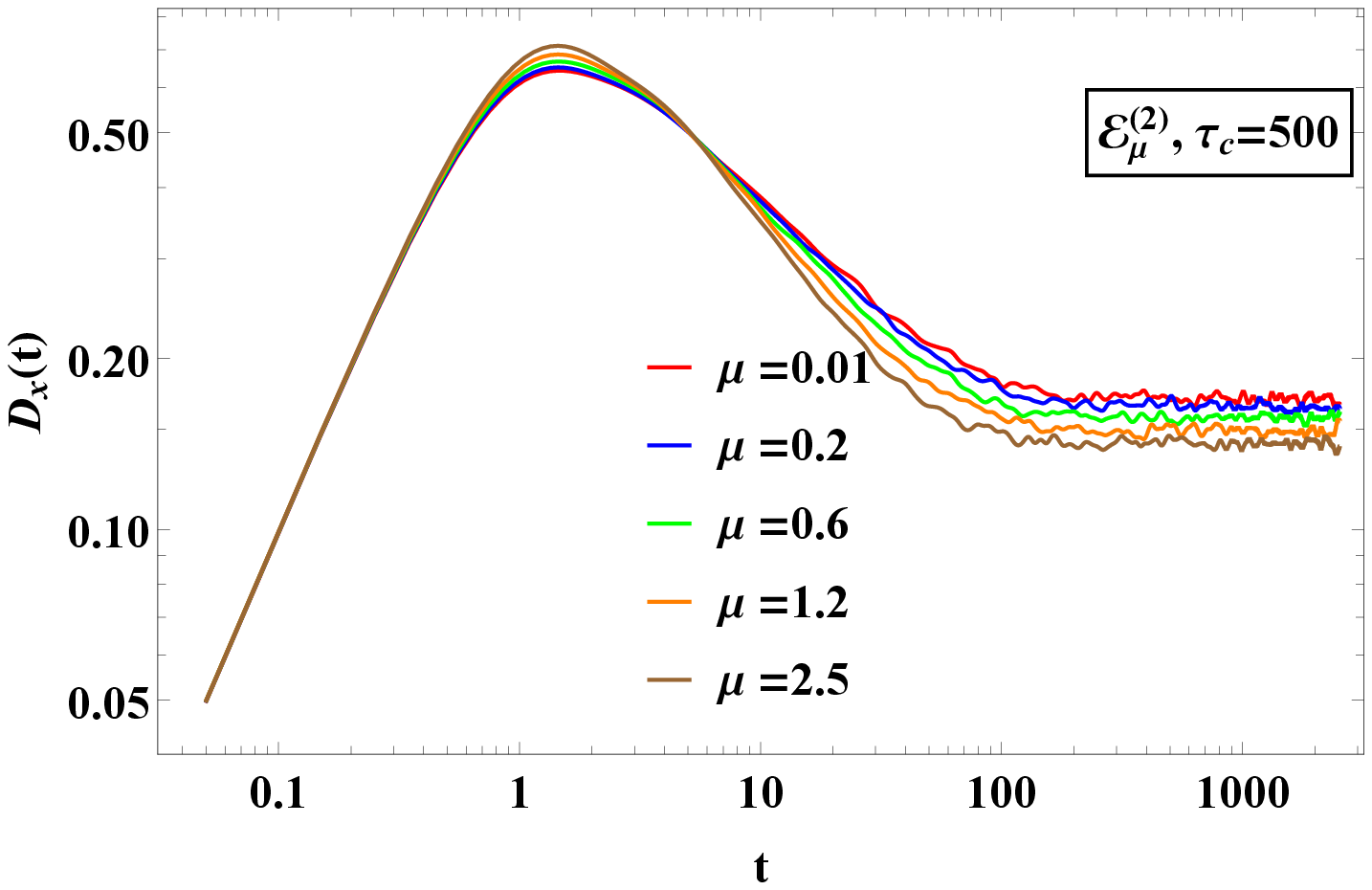}%
	}
	\subfloat[\label{fig_6c}]{
	\includegraphics[width=.33\linewidth]{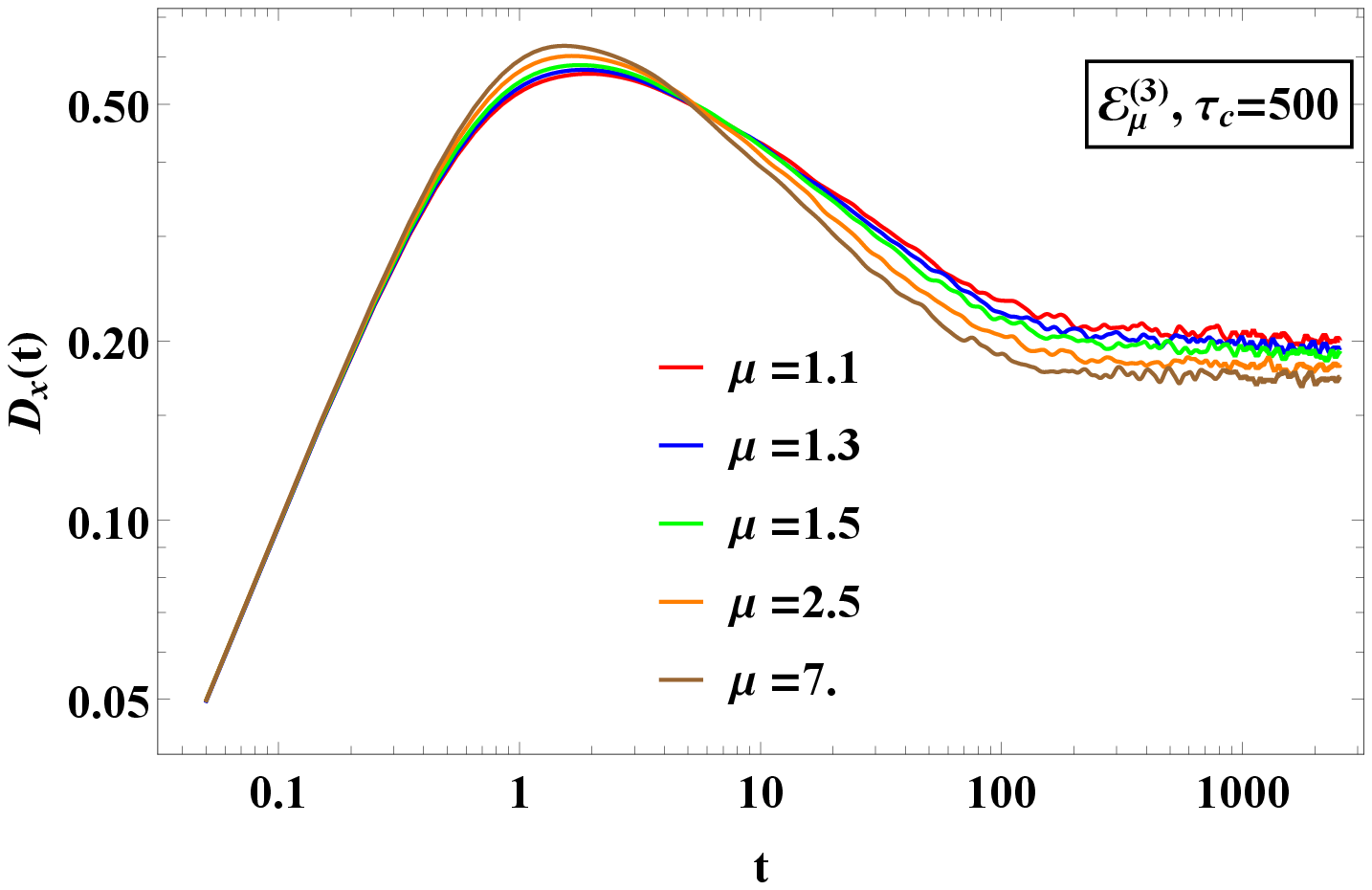}}
	\caption{Running diffusion profiles for all correlations $\mathcal{E}^{(1-3)}$ at different $\mu,\nu$ values at $\tau_c=500$.}
\end{figure}

\begin{figure}
	\subfloat[\label{fig_6d}]{
		\includegraphics[width=.8\linewidth]{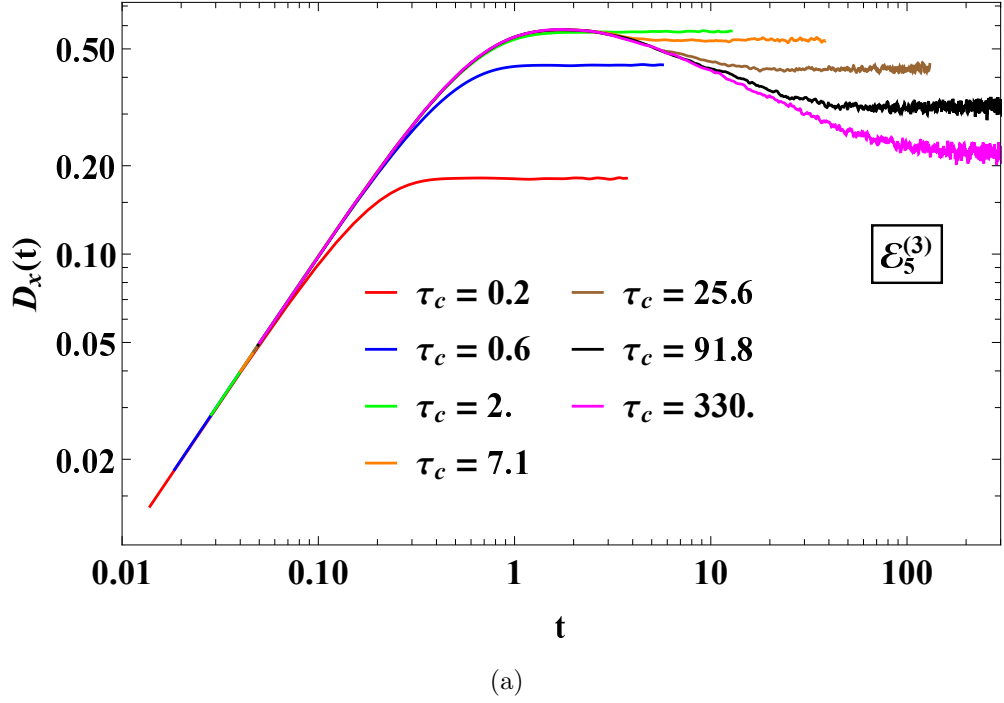}%
	}
	\caption{Running diffusion profiles for different $\tau_c$ values in the case of $\mathcal{E}^{(3)}_5$.}
\end{figure}

We consider now the more complex case of time-dependent turbulent fields, when $\tau_c = $ finite. The time-decorrelation leads to pure diffusive features, i.e. $\lim\limits_{t\to\infty}D_x(t) = $ finite. The time-dependent $D_x(t)$ can be seen in Figs. \eqref{fig_6a}-\eqref{fig_6c} for all three correlations $\mathcal{E}^{(1-3)}$ at different values of the parameters $\mu,\nu$ and setting $\lambda_c=1$. All profiles resemble the generic data shown in Fig. \eqref{fig_2a}. The case $\tau_c = 500$ was chosen as representative for the strong turbulent (percolative) regime. In practice, a large span of differential correlations times $T_D=\tau_c\in (0.15,10^3)$ were simulated. We show in Fig. \eqref{fig_6d} the diffusion coefficients $D_x(t)$ for the same correlation $\mathcal{E}_5^{(3)}$ but for different $\tau_c$ values. At small times $t\ll t_{peak}$ the diffusion is linear $D_x(t)=V^2t^1$ and independent of $\tau_c$, while at larger times it saturates after roughly $t\sim T_I$. 

In many systems of interest (fusion plasmas, astrophysical magnetic fields, etc.) there is a clear separation of scales between: the time scale $\sim T_I$ of turbulent transport is much smaller than macroscopic, experimentally relevant, time-scales of dynamics. Because of that, only the asymptotic diffusion $D_x^\infty$ is of practical relevance. Consequently, we shall focus on $D_x^\infty$ from now on.

\begin{figure}
	\subfloat[\label{fig_8a}]{
		\includegraphics[width=.33\linewidth]{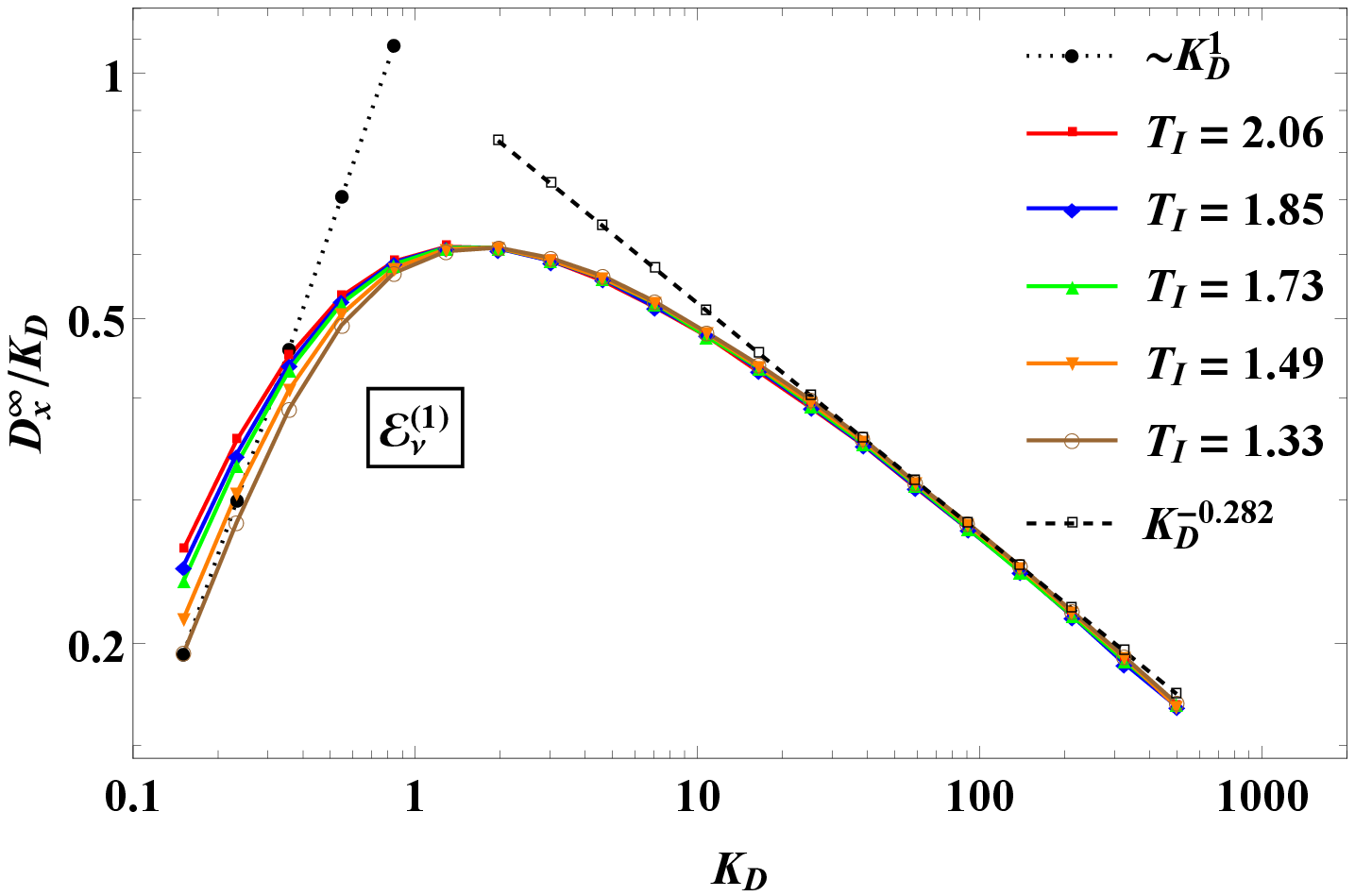}%
	}
	\subfloat[\label{fig_8b}]{
		\includegraphics[width=.33\linewidth]{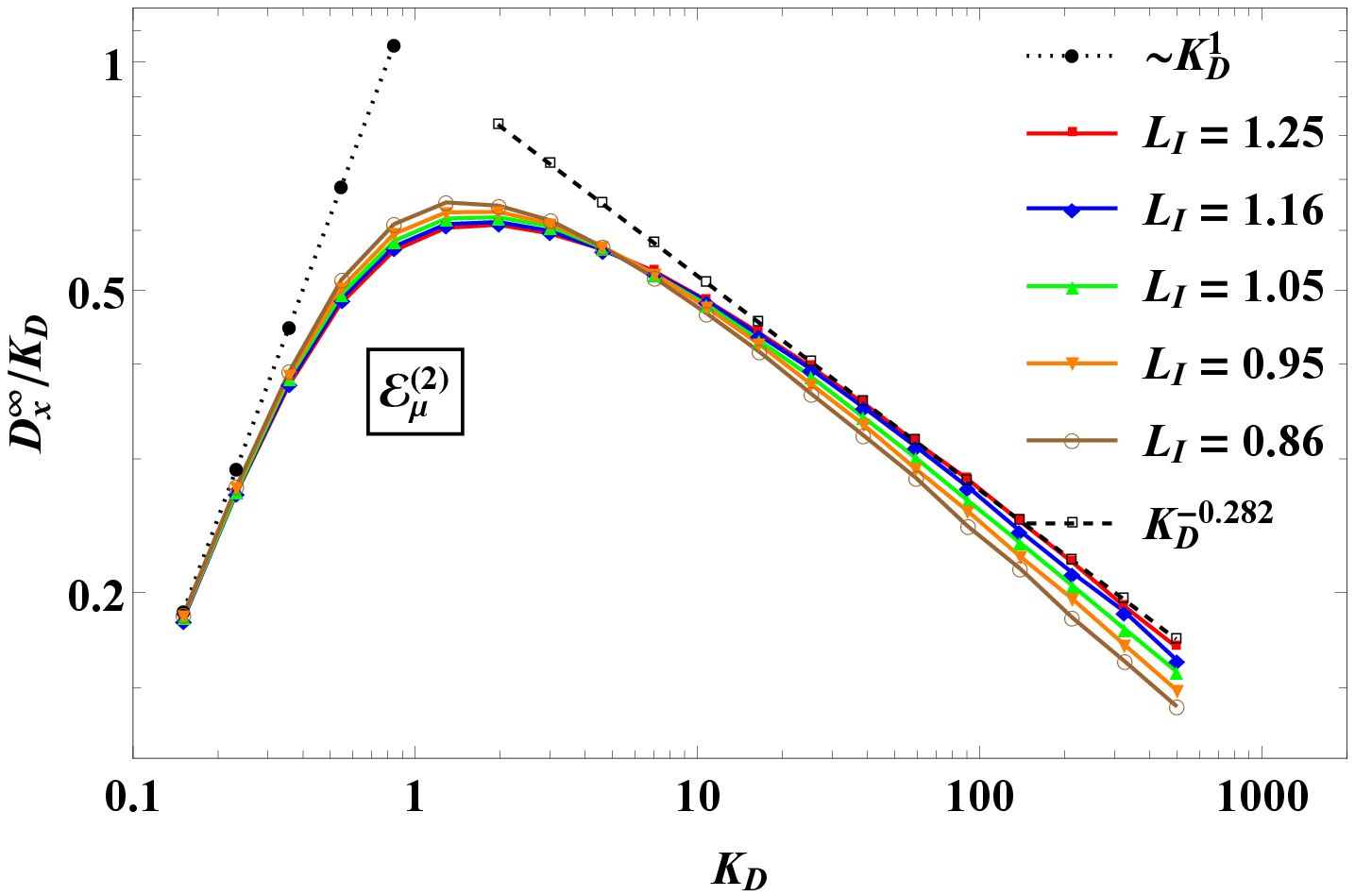}%
	}
	\subfloat[\label{fig_8c}]{
		\includegraphics[width=.33\linewidth]{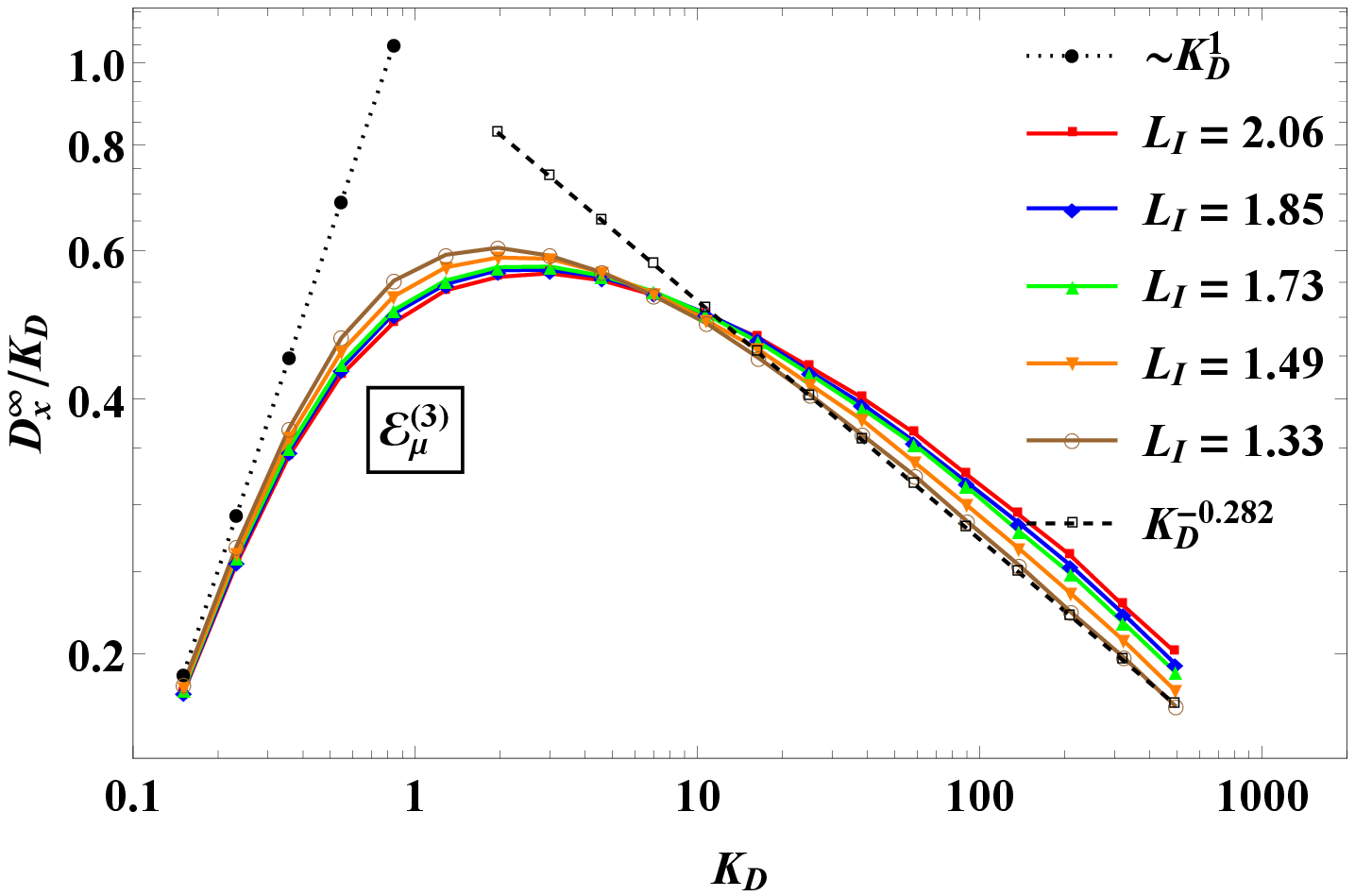}%
	}
	\caption{Asymptotic diffusion coefficients $D^\infty_x$ at various differential Kubo numbers $K_D$ and integral correlation lengths/times $L_I,T_I$}
\end{figure}

In Figs. \eqref{fig_8a}-\eqref{fig_8c} we show how $D_x^\infty/K_D$ is dependent on the differential Kubo number $K_D$, for all three classes of correlations. The quasilinear behaviour $D_x^\infty = V^2T_I$ can be observed in the region $K_D\ll 1$ for all cases. A maximum of $D_x^\infty/K_D$ is present at $K_D\approx T_I/T_D$, then the profile decays algebraically in the strong turbulent regime $K_D\gg 1$. 

Regarding the correlation $\mathcal{E}^{(1)}$ it is interesting to note that the $D_x^\infty$ profiles are different only in the quasilinear regime, while beyond $K_D>1$ become identical. This suggests that in strong turbulence the specific details of time-decorrelation are not important, only those of space correlations.

Regarding $\mathcal{E}^{(2)}$, it appears that the space oscillations of the correlation function are reflected in the $D_x^\infty$ profile in a qualitatively similar manner with the frozen $D_x(t)$ (Fig. \eqref{fig_3a}. The position of the maxima is roughly stable to correlation changes, but its value it is not. Moreover, in the strong regime, $K_D\gg 1$, the slope is dependent on $L_I$ (or, equivalently, $\mu$). Similar observations can be drawn for the case of $\mathcal{E}^{(3)}$ which mimics the frozen time profiles.

\begin{figure}
	\subfloat[\label{fig_9a}]{
		\includegraphics[width=.48\linewidth]{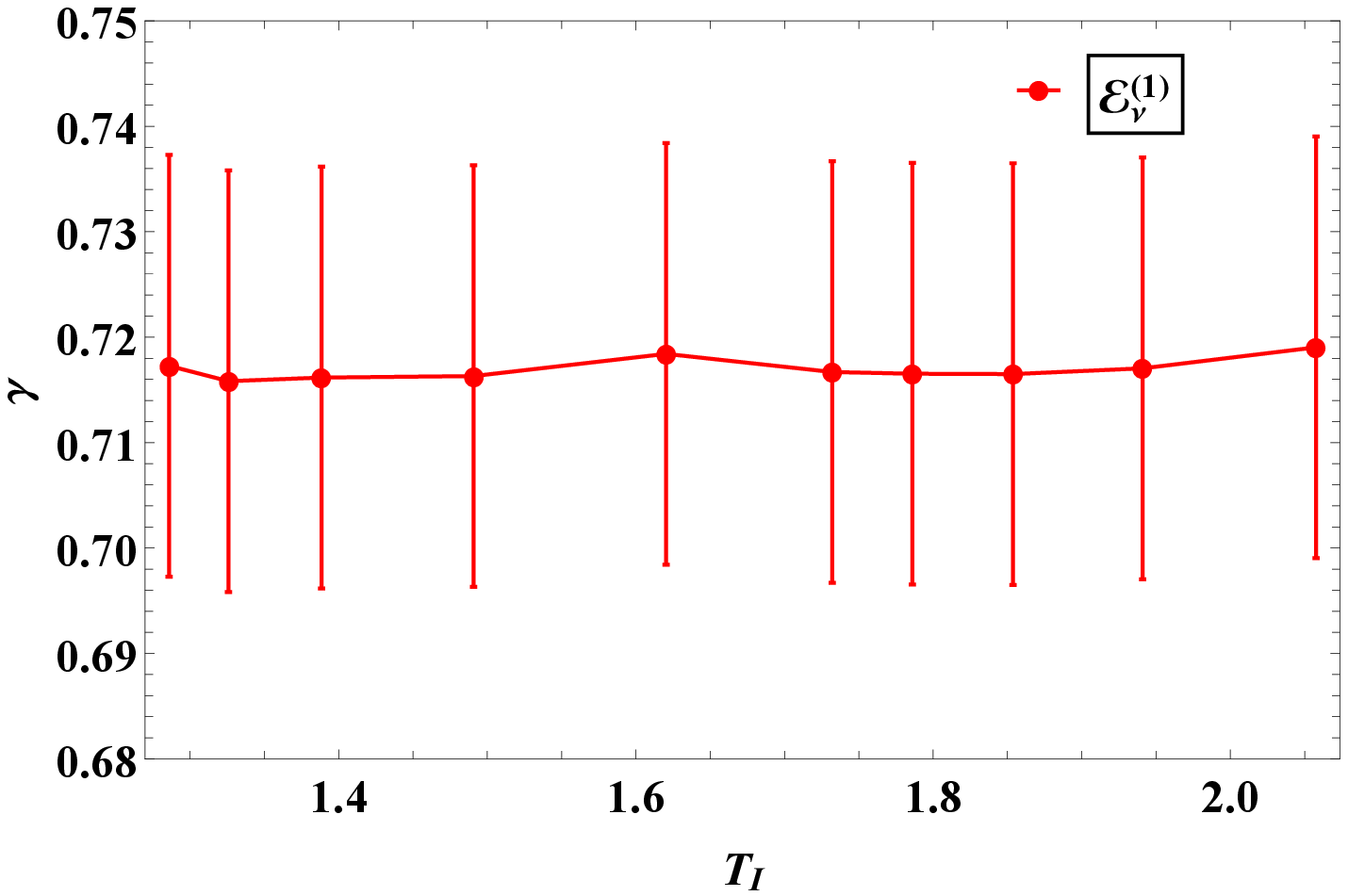}%
	}\hspace*{0.1cm}
	\subfloat[\label{fig_9b}]{
		\includegraphics[width=.48\linewidth]{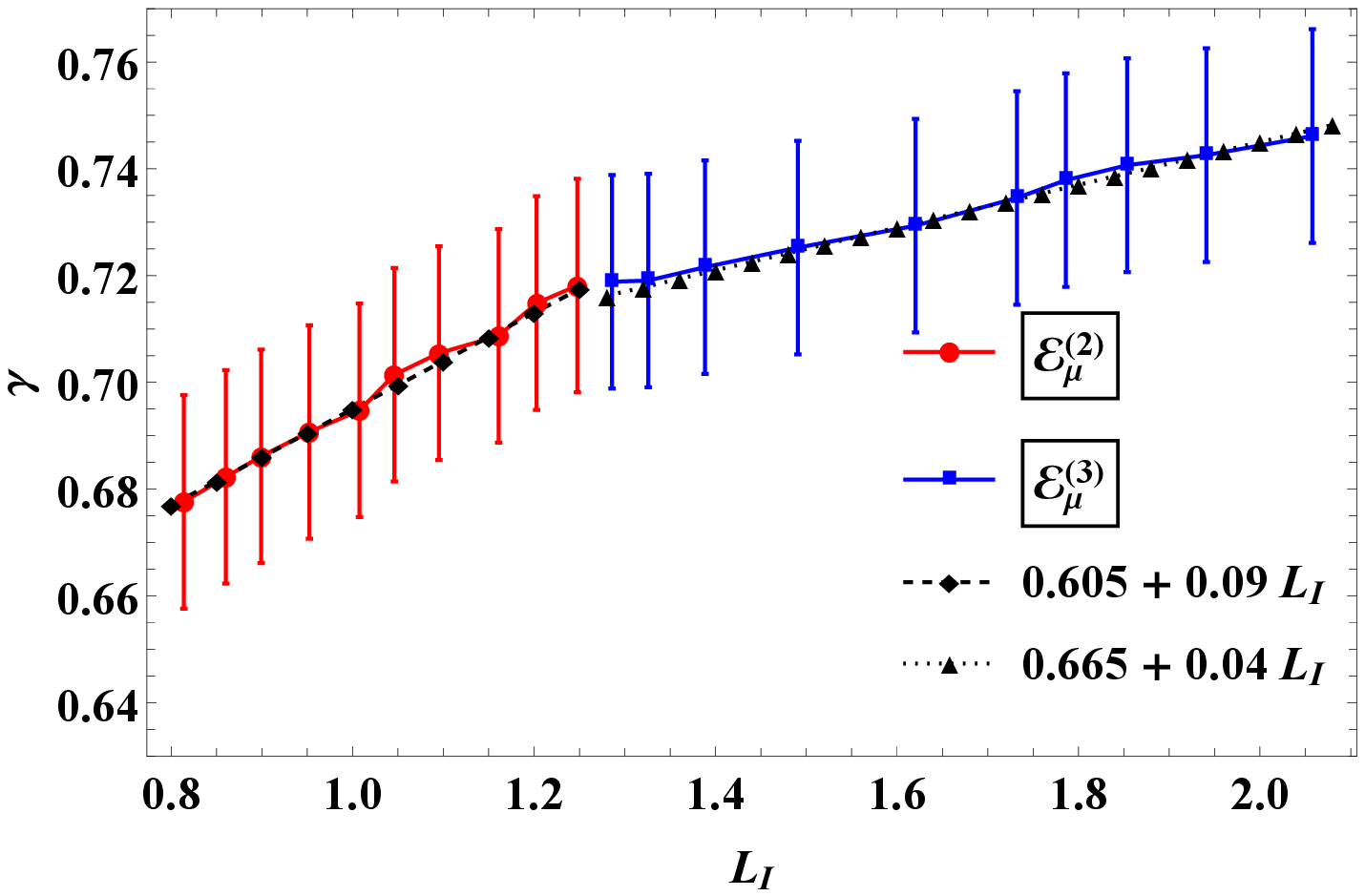}%
	}
	\caption{The dependence of the anomalous exponent $\gamma$ with the times-space integral scales. First figure is for $\mathcal{E}^{(1)}$ and the second for $\mathcal{E}^{(2-3)}$.}
\end{figure}

\begin{figure}
		\includegraphics[width=.6\linewidth]{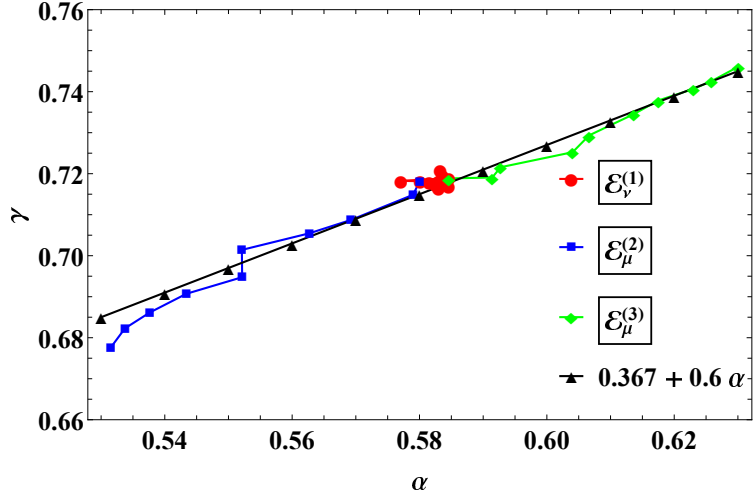}
		\caption{The dependence between the anomalous exponents $\gamma$ from the time-decorrelated case and the exponent $\alpha$ from the frozen case for all correlation classes $\mathcal{E}^{(1-3)}$.}
\label{fig_10}
\end{figure}

Our results confirm the algebraic behaviour of the asymptotic diffusion (percolative) with the differential Kubo number as $D_x^\infty = A K_D^\gamma$. Moreover, analyzing the numerical data, we have found that the proportionality constant is exactly $A = \lambda_c^2/\tau_c = L_D^2/T_D$. Using the exact quasilinear result $D_x^\infty = V^2 T_I$ we can provide a gross estimation for diffusion profiles in all regimes: 

$$D_x^\infty (K_D) \approx \frac{L_D^2}{T_D}\frac{T_IK_D^2}{T_D+T_IK_D^{-\gamma+2}}$$

We explore further the dependence (universality) of the $\gamma$ exponent on turbulence features, in particular, the integral scale length $L_I$ (eq. \eqref{eq_6}). We plot the results in Figs. \eqref{fig_9a}-\eqref{fig_9b}, assisted by error bars that indicate the numerical error due to statistical fluctuations and interpolation procedures. The qualitative observation from Fig. \eqref{fig_8a} that $\mathcal{E}^{(1)}_\nu$ has no impact on the slope of $D_x^\infty(K_D)$ at large $K_D$ is confirmed in Fig. \eqref{fig_9a} where the value $\gamma\approx 0.718$, very close to Isichenko's value $\gamma = 7/10$, has been found for all $T_I$.

The dependence of $\gamma$ with the integral scale $L_I$ shown in Fig. \eqref{fig_9b} reveals a coherent monotonically dependence $\gamma(L_I)$. For both classes of space correlations $\mathcal{E}^{(2-3)}$, the dependence is linear, but with different slopes, similar with the dependence of the $\alpha(L_I)$ exponent.

It is worth mentioning that the dependence of various quantities $\alpha, t_{peak}, \gamma$ (Figs. \eqref{fig_4a},\eqref{4c} and \eqref{fig_9b}) with the integral scale length $L_I$ is not coherent between $\mathcal{E}^{(2)}$ and $\mathcal{E}^{(3)}$ (different slopes). This suggests that $L_I$ is not the best universal measure of turbulent features and something more intricate should be searched for. This could be the purpose of a future work. 

Finally, we can ask for the relation between asymptotic diffusion in the decorrelated case, $\tau_c = $ finite, and the percolative limit, $\tau_c = \infty$. If an analytical relation would be available, that could enable the evaluation of diffusion coefficients in a large range of cases solely from the frozen profile. Unfortunately, we have not been able to find any simple relation between $D_x(t\to\infty;\tau_c)$ and $D_x(t;\tau_c\to\infty)$. The only observation, important as it is, can be seen in Fig. \eqref{fig_10} where we plot the $\gamma$ exponent for the strong turbulent regime in the time-decorrelated case vs. the $\alpha$ exponent for the long-time behaviour of diffusion in the frozen turbulent limit, i.e. $D_x^\infty\sim K_D^\gamma, D_x(t)\sim t^{\alpha-1}$. A linear dependence is found: $\gamma \approx 0.367+0.6\alpha$. This implies that we can estimate $D_x^\infty$ as:

\begin{align}
	D_x^\infty(K_D) &= D_x^{frozen}\left(t = T_DK_D^{-\alpha/\gamma}\right)\\
	\gamma &= 0.367+0.6\alpha.
\end{align}

\section{Conclusions}
\label{section_4}

In the present work we have investigated the features of diffusive transport in two-dimensional incompressible turbulent fields. The turbulence is described by its spectrum, or, equivalently, its correlation function. In order to encompass as general features as possible, we have considered three distinct classes of correlations with: long-range time dependencies $\mathcal{E}^{(1)}$, space oscillations  $\mathcal{E}^{(2))}$, or long-range space dependencies $\mathcal{E}^{(3)}$. Employing the method of direct numerical simulations, the transport coefficients have been computed on a large span of physical regimes. We have used unprecedented numerical resolution in order to capture in great detail any small effects induced by the geometrical attributes of turbulence. 

For the frozen regime of time-independent turbulence we have numerically confirmed that the diffusion has a linear behaviour at small times, while being anomalous ($\sim t^{\alpha-1}$) in the long-time limit. We have found that both the maximal diffusion $D_{peak}$ and the slope of the algebraic decay $\alpha$ are dependent on the correlation function. The analytical and approximate results for the frozen regime are summarized in Table \eqref{table_1}.

\begin{table}
	\begin{tabular}{ |c|c|c|c|} 
		\hline
		 percolative   & $D_x(t)$ & $D_{peak}/(L_IV)$ & $\alpha$\\
		\hline
 	 $\tau_c = \infty$ & $V^2 t^1,  t\ll t_{peak}$ & $1.63-1.27L_I+0.3L_I^2$ & $0.423+0.128L_I$, for $\mathcal{E}_\mu^{(2)}$\\
\hline  
		  & $\lambda_c^{2-\alpha}V^{\alpha}t^{\alpha-1}, t\gg t_{peak} $ & $1.63-1.27L_I+0.3L_I^2$ & $0.502+0.065L_I$, for $\mathcal{E}_\mu^{(3)}$\\
		  \hline
	\end{tabular}
\caption{Summary of analytical and approximate results for the case of frozen turbulence.}
\label{table_1}
\end{table}

For the time decorrelated case we have confirmed that diffusion is roughly $V^2 t^1$ at small times, then saturates to finite values at longer times. The dependence of asymptotic diffusion $D_x^\infty$ on Kubo numbers indicates that in the quasilinear case $D_x^\infty = V^2 T_I$. In the percolative regime, the dependence of diffusion on $K_D$ is algebraic $K_D^{\gamma}$. The critical exponent $\gamma$ is insensitive to the features of time decorrelation but is linearly dependent on the integral space scale. Furthermore, $\gamma$ increases linearly with the $\alpha$ exponent. The analytical and approximate results for this case are summarized in Table \eqref{table_2}.

\begin{table}
	\begin{tabular}{ |c|c|c|c|c|} 
		\hline
	decorrelated     & $t\ll t_{peak}$  & $D_x^\infty, K_\star\ll 1$ & $D_x^\infty, K_\star\gg 1$ & $\gamma$\\
		\hline
	$\tau_c=$ finite &  $D_x(t) = V^2 t^1$ & $ V^2 T_I = L_I^2/T_I K_I^2$ & $L_D^2/T_DK_D^{\gamma}$ &  $0.665+0.04L_I$, for $\mathcal{E}_\mu^{(3)}$\\ 
		\hline
	   &    & $ V^2 T_I = L_D^2/T_D^2 T_I  K_D^2 $ &   & $0.605+0.09L_I$, for $\mathcal{E}_\mu^{(2)}$\\ 
\hline
	\end{tabular}
\caption{Summary of analytical and approximate results for the decorrelated regime.}
\label{table_2}
\end{table}

Finally, simple approximate, analytical, formulas for the relation between diffusion and Kubo numbers are shown here for practical purposes.

\begin{align}
	\label{interpol}
	D_x^\infty(K_D) &= \frac{L_D^2}{T_D}\frac{\left(\frac{T_I}{T_D}\right)K_D^2}{1+\left(\frac{T_I}{T_D}\right)K_D^{2+\gamma}}\\	D_x^\infty(K_I) &= \frac{L_I^2}{T_I}\frac{K_I^2}{1+\left(\frac{T_D}{T_I}\right)^{\gamma+1}\left(\frac{L_I}{L_D}\right)^{\gamma+2}K_I^{2+\gamma}}.	
\end{align}

We acknowledge that the linear interpolation of critical exponents $\alpha,\gamma$ as functions of $L_I$ (Figs. \eqref{fig_4c}, \eqref{fig_9b}) is not fully coherent between long-range and oscillating correlation functions $\mathcal{E}^{(3)},\mathcal{E}^{(2)}$. This suggests that a more appropriate global measure of Eulerian turbulent spectra is possible, beyond integral scales $L_I$, and are to be sought in future works.

\section*{Acknowledgements}
This research was partially supported by Romanian Ministry of Research, Innovation and Digitalization under Romanian National Core Program LAPLAS VII – contract no. 30N/2023.

\bibliographystyle{apsrev}
\bibliography{biblio}


\end{document}